\documentclass[preprint2]{aastex}

\usepackage{rotating}

\newcommand{\degree}{\ensuremath{^{\circ}}}

\newcommand{\moo}{\rm $\mu$m}

\shorttitle{New MFIR and radio observation of the 2Jy sample}
\shortauthors{Dicken et al.}

\begin{document}

\title{The origin of the infrared emission in radio galaxies I: \\ new mid- to far-infrared and radio observations of the 2Jy sample}

\author{D. Dicken\altaffilmark{1}, C. Tadhunter\altaffilmark{1}, R.
Morganti\altaffilmark{2,3}, C. Buchanan\altaffilmark{4}, T. Oosterloo\altaffilmark{2,3}, 
D. Axon\altaffilmark{5} }

\altaffiltext{1}{Department of Physics and Astronomy, University of
    Sheffield, Hounsfield Road, Sheffield, S3 7RH;
    d.dicken@sheffield.ac.uk, c.tadhunter@sheffield.ac.uk}
\altaffiltext{2}{ASTRON, P.O. Box 2,
    7990 AA Dwingeloo, Netherlands; morganti@astron.nl, oosterloo@astron.nl}
\altaffiltext{3}{Kapteyn Astronomical Institute, University of Groningen Postbus 800, 9700 AV Groningen, Netherlands}
\altaffiltext{4}{School of Physics, University of Melbourne, Victoria 3010, Australia; clb@physics.unimelb.edu.au}
\altaffiltext{5}{Department of Physics and Astronomy, Rochester Institute of Technology, 84 Lomb Memorial Drive, Rochester NY 14623; djasps@rit.edu}

\begin{abstract}
As part of a large study to investigate the nature of the longer
wavelength continuum emission of radio-loud AGN, we present new mid to
far-infrared (MFIR) and high frequency radio observations for a
complete sample of 2Jy powerful, southern radio galaxies at
intermediate redshifts ($0.05<z<0.7$). Utilizing the sensitivity of
the {\it Spitzer} Space Telescope, we have made deep {\it MIPS} observations at
the wavelengths of 24, 70 and 160\moo, detecting 100\% of our sample
at 24\moo\, and 90\% at 70\moo. This high detection rate at MFIR
wavelengths is unparallelled in samples of intermediate redshift radio
galaxies. Complementing these results, we also present new high
frequency observations (15 to 24Ghz) from the Australia Telescope
Compact Array ({\it ATCA}) and the Very Large Array ({\it VLA}), which are used to
investigate the potential contamination of the MFIR continuum by
non-thermal synchrotron emission. With the latter observations we
detect the compact cores in 59\% of our complete sample, and deduce
that non-thermal contamination of the MFIR continuum is significant in
a maximum of 30\% of our total sample. MFIR fluxes, radio fluxes and
spectral energy distributions for the complete sample are presented
here, while in a second paper we will analyse these data and discuss
the implications for our understanding of the heating mechanism for
the warm/cool dust, star formation in the host galaxies, and the
unified schemes for powerful radio sources.
\end{abstract}


\section{Introduction}
\label{sec:intro}

The mid to far-infrared (MFIR) is one of the most profitable
wavelength ranges for investigating AGN and associated phenomena. This
is primarily because AGN emission suffers less from effects of
obscuration at these wavelengths. Moreover, the thermal infrared
samples the emission that is absorbed and then re-radiated by dust,
indirectly sampling the radiative power of the active core and any
starburst component that may be present. Therefore this wavelength
range is crucial both for testing the unified schemes for AGN
(e.g. \citealp{barthel89}), and investigating the link between AGN and
starburst activity as part of the evolution of the host galaxies
(\citealp{rowan95}, \citealp{haas98}, \citealp{archibald01}).

In terms of testing unified schemes, radio-loud samples of AGN have
the advantage that their extended radio emission is isotropic,
allowing samples to be selected free from the orientation bias that
affects samples selected at X-ray, optical and near-IR
wavelengths. However, in order to fully realize the potential of MFIR
observations for studies of radio galaxies it is important to obtain
observations of carefully selected samples that attain a high level of
completeness in terms of detections.  Unfortunately, observing in the
MFIR is technically more challenging than at near-infrared
wavelengths.  Therefore, although the {\it IRAS} observatory made many
significant contributions to our understanding of the MFIR infrared
properties of radio-loud AGN (\citealp{neugebauer86},
\citealp{golombek88}, \citealp{knapp90}, \citealp{impey93},
\citealp{heckman92,heckman94} \citealp{hes95}), it had a relatively
low sensitivity, resulting in a typical MFIR detection rate of less
than 30\%, even for radio galaxies in the local universe ($z<0.3$).

Subsequent attempts were made to observe samples of radio galaxies
with the {\it ISO} observatory (\citealp{haas98,haas04},
\citealp{polletta00}, \citealp{bemmel00},
\citealp{meisenheimer01}). However, meager, sometimes heterogeneous
samples, and the modest sensitivity of the observatory, only allowed
for a small improvement on the achievements of {\it IRAS}. Studies of
samples of radio-loud AGN with {\it ISO} have typical detection rates of no
more than 50\%\ in the far-IR, impeding attempts to understand the
MFIR emission.

The launch of {\it Spitzer} in 2003, brought an opportunity to make
observations with orders of magnitude improved sensitivity at MFIR
wavelengths. However, published {\it Spitzer} studies of radio loud
AGN are either based on heterogeneous samples (e.g. \citealp{shi05}),
or on samples that are relatively faint and distant, resulting in a
low detections rates at 70\moo\ (e.g. \citealp{cleary07},
\citealp{seymour07}).

A further problem is that many previous MFIR studies of radio-loud AGN
samples are selected from the 3C radio catalogue, which are hampered by
the lack of published high quality optical spectroscopic observations
for many of the objects. In contrast, the southern 2Jy sample
\citep{tadhunter93} is unique in the sense that deep spectra have been
published for the whole sample
(\citealp{tadhunter93,tadhunter98,tadhunter02}; \citealp{wills02},
\citealp{holt07}). Measured emission line luminosities for this sample
provide information about the intrinsic power of the AGN
\citep{tadhunter98}. Moreover, careful modelling of the continuum
spectra provides key information about the stellar populations in the
host galaxies and, in particular, in the presence of young stellar
populations (\citealt{tadhunter02}; \citealt{wills04}).  The 2Jy
sample has also been extensively observed at radio wavelengths
(\citealp{morganti93,morganti97,morganti99}), allowing links to be
made between radio and optical properties and independent estimates of
the orientation of the sources to the line of sight. In view of its
completeness and the availability of deep spectroscopic and radio
data, which makes it well suited to investigating the nature of the
MFIR emission and testing the unified schemes, we have undertaken a
program of deep imaging with {\it Spitzer/MIPS} of the 2Jy sample.

Given that we are interested in the thermal emission and heating
mechanism of the warm/cool dust sampled through the MFIR fluxes,
possible contamination from non-thermal synchrotron sources could
potentially bias our conclusion. Therefore, to complement the MFIR
observations, we have also undertaken high frequency observations at
radio wavelengths (15 to 22 GHz) using {\it ATCA} and the {\it VLA}, with the aim
of investigating the potential degree of non-thermal contamination.

A preliminary analysis of the results from this program was presented in
\citet{tadhunter07}. Here we present the MFIR and radio core data, and
discuss the potential non-thermal contamination of the MFIR
continuum. A second paper comprises an in-depth analysis of these
data (Dicken et al. 2008, in preparation).

\section{Sample selection}
\label{sec:sample}
The sample for this study comprises a complete sub-sample of all 46
steep-spectrum powerful radio galaxies and quasars with redshifts 0.05
$<$ z $<$ 0.7, and flux densities $S_{2.7GHz}>$ 2Jy from the sample of
\citet{tadhunter93}. We define the the steep spectrum selection as
$\alpha^{4.8}_{2.7} > 0.5$ ($F_{\nu} \propto \nu^{-\alpha}$), which
excludes 16 objects within the redshift range. We also include
\object{PKS0347$+$05}, which has since proved to fulfil the same
selection criteria \citep{diserego94}. The lower redshift limit has
been set to ensure that these galaxies are genuinely powerful sources,
and the steep spectrum selection for the quasars rules out objects
dominated by emission from the beamed relativistic jet and core
components. Overall, the full sample of 46 objects included a mixture
of broad line radio galaxies/radio loud quasars (BLRG/Q: 35\%),
narrow line radio galaxies (NLRG: 43\%) and weak line radio galaxies
(WLRG: 22\%). In terms of radio morphology classification, our
complete sample includes 72\% FRII sources, 13\% FRI sources and 15\%
compact steep spectrum/gigahertz peak spectrum objects.

For comparison purposes we also observed or collected data from the
{\it Spitzer/IRAS} archive, for the flat spectrum, core dominated
objects \object{3C273}, \object{PKS0521$-$36} and
\object{PKS1549$-$79}. Note that \object{PKS1549$-$79} is a
particularly interesting source, because it a rare example of a
powerful, flat spectrum radio source that is classified as a galaxy at
optical wavelengths; the nature of this object is discussed in detail
in \citet{holt06}. However, these objects are not part of what we
refer to as our complete sample. Our complete sample of 46 objects
comprises the 49 objects with {\it Spitzer} or {\it IRAS} observations
presented below, minus the 3 comparison flat spectrum objects.

\section{MFIR observations and reduction}
\label{sec:infrared}
\subsection{Observations}
We have obtained new observations for 43 objects using the {\it Spitzer}
Space Telescope and, for 37 of these, we have obtained the very first
MFIR detections. In addition, we include results from data
obtained for 5 other targets from the {\it Spitzer} archive Reserved
Observations Catalogue (ROC): \object{PKS0915-11} (3C218),
\object{PKS1226$+$02} (3C273), \object{PKS1559$+$02} (3C327),
\object{PKS1648$+$05} (3C348) and \object{PKS2221-02} (3C445). The
observations were carried out between August 2005 and Jan 2007, and
the ROC targets were observed between March and November 2004.

All targets were observed with the Multiband Imaging Photometer for
{\it Spitzer} ({\it MIPS}: \citealp{rieke04}) at wavelengths of 24, 70
and 160\moo, apart from 5 targets that were not observed at 160\moo\
due to scheduling reasons. Details of the observations are shown in
Table \ref{tbl-1}.

\subsection{Reduction}

The {\it MIPS} instrument on board {\it Spitzer} takes short exposure images in a
dither pattern. These are then combined to make a final `mosaiked'
image. Rather than use the pipeline processed data, we have re-reduced
data for all the 48 objects in the sample starting from the Basic
Calibrated Data (BCD) files to produce final mosaiked images for
analysis. The reduction was carried out at Rochester Institute of
Technology and The University of Sheffield using the MOPEX software
reduction package \citep{makovoz06} provided by the {\it Spitzer} Science
Center (SSC). Due to the different nature of the detectors and
sensitivities at the 3 wavelengths observed, the reduction method we
have selected is different for each. Thus we discuss the 3 bands
separately below.

\begin{figure*}
\epsscale{1.5}
\plotone{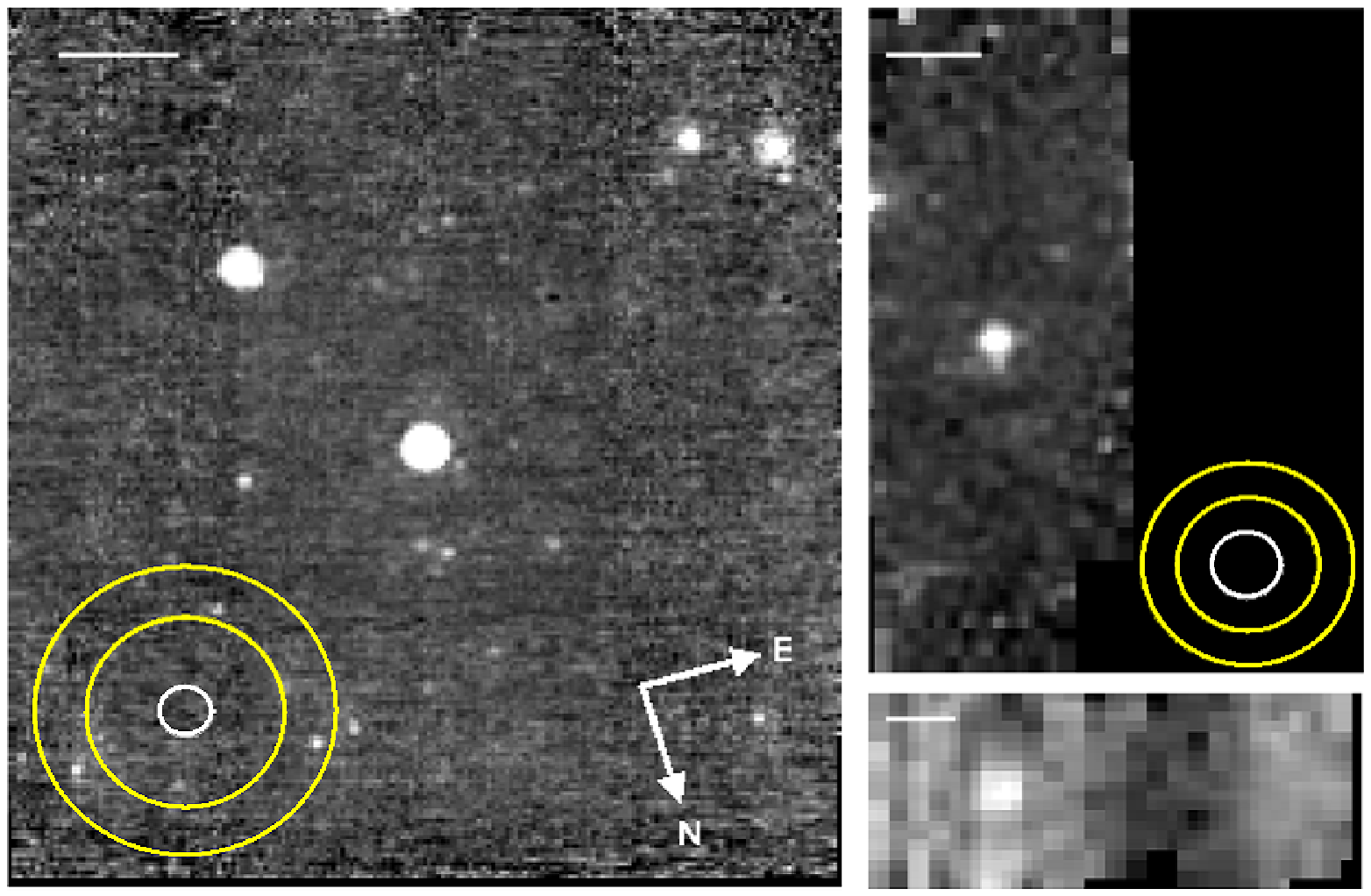}
\caption{Example of a {\it MIPS} mosaiked images of PKS1814$-$63. Left:
24\moo\ image with an example aperture. Top Right: 70\moo\ image with
an example aperture. Bottom Right: 160\moo\ image. The white circles
indicate the apertures used for the object flux measurements, while
the yellow circles delimit the annuli used to estimate the background
sky flux. The bar in the top right hand corner of each image is of
length 1 arcmin. \label{1814}}
\end{figure*}

\begin{deluxetable}{c@{\hspace{-2mm}}l@{\hspace{-2mm}}c@{\hspace{-2mm}}r@{\hspace{-3mm}}c@{\hspace{-3mm}}c@{\hspace{-1mm}}c@{\hspace{-1mm}}c@{\hspace{-3mm}}r@{\hspace{-3mm}}r@{\hspace{-3mm}}r}
\tabletypesize{\scriptsize}
\tablecaption{{\it Spitzer} Sample and Observational Data \label{tbl-1}}
\tablewidth{0pt}
\tablehead{
\colhead{PKS Name}{\hspace{-2mm}} & \colhead{Other}{\hspace{-2mm}} & \colhead{RA(J2000)}{\hspace{-2mm}} &  \colhead{Dec(J2000)}{\hspace{-3mm}} &\colhead{Opt. Class}{\hspace{-3mm}} & \colhead{Rad. Class}{\hspace{-1mm}} &
\colhead{z}{\hspace{-1mm}} & \colhead{Obs. Date}{\hspace{-3mm}} & \colhead{$t_{int}$24(s)}{\hspace{-3mm}} & \colhead{$t_{int}$70(s)}{\hspace{-3mm}} & \colhead{$t_{int}$160(s)}
}
\startdata
0023$-$26	&	\phantom{a}		&	00 25 49.18	&	-26 02 12.8	\phantom{aa}	&	NLRG	&	CSS	&	0.322	&	Dec-05	&	180.4	\phantom{aa}	&	543.3	\phantom{aa}	&	167.8	\phantom{aa}	\\
0034$-$01	&	\phantom{a}	3C015	&	00 37 04.14	&	 -01 09 08.2	\phantom{aa}	&	WLRG	&	FRII	&	0.073	&	Jan-06	&	180.4	\phantom{aa}	&	543.3	\phantom{aa}	&	167.8	\phantom{aa}	\\
0035$-$02	&	\phantom{a}	3C17	&	00 38 20.51	&	-02 07 40.1 	\phantom{aa}	&	BLRG	&	(FRII)	&	0.220	&	Jul-06	&	180.4	\phantom{aa}	&	543.3	\phantom{aa}	&	167.8	\phantom{aa}	\\
0038$+$09	&	\phantom{a}	3C18	&	00 40 50.53	&	+10 03 26.8 	\phantom{aa}	&	BLRG	&	FRII	&	0.188	&	Jan-06	&	180.4	\phantom{aa}	&	543.3	\phantom{aa}	&	167.8	\phantom{aa}	\\
0039$-$44	&	\phantom{a}		&	00 42 09.03	&	-44 14 01.3 	\phantom{aa}	&	NLRG	&	FRII	&	0.346	&	Dec-05	&	180.4	\phantom{aa}	&	543.3	\phantom{aa}	&	167.8	\phantom{aa}	\\
0043$-$42	&	\phantom{a}		&	00 46 17.75	&	-42 07 51.4 	\phantom{aa}	&	WLRG	&	FRII	&	0.116	&	Dec-05	&	180.4	\phantom{aa}	&	543.3	\phantom{aa}	&	167.8	\phantom{aa}	\\
0105$-$16	&	\phantom{a}	3C32	&	01 08 16.90	&	 -16 04 20.6 	\phantom{aa}	&	NLRG	&	FRII	&	0.400	&	Jul-06	&	180.4	\phantom{aa}	&	543.3	\phantom{aa}	&	167.8	\phantom{aa}	\\
0117$-$15	&	\phantom{a}	3C38	&	01 20 27.16	&	 -15 20 17.8 	\phantom{aa}	&	NLRG	&	FRII	&	0.565	&	Jul-06	&	180.4	\phantom{aa}	&	543.3	\phantom{aa}	&	167.8	\phantom{aa}	\\
0213$-$13	&	\phantom{a}	3C62	&	02 15 37.50	&	 -12 59 30.5 	\phantom{aa}	&	NLRG	&	FRII	&	0.147	&	Jan-07	&	180.4	\phantom{aa}	&	545.3	\phantom{aa}	&	167.8	\phantom{aa}	\\
0235$-$19	&	\phantom{a}	OD-159	&	02 37 43.45	&	 -19 32 33.3 	\phantom{aa}	&	BLRG	&	FRII	&	0.620	&	Feb-06	&	180.4	\phantom{aa}	&	545.3	\phantom{aa}	&	167.8	\phantom{aa}	\\
0252$-$71	&	\phantom{a}		&	 02 52 46.26	&	-71 04 35.9 	\phantom{aa}	&	NLRG	&	CSS	&	0.566	&	Aug-05	&	180.4	\phantom{aa}	&	545.3	\phantom{aa}	&	167.8	\phantom{aa}	\\
0347$+$05	&	\phantom{a}		&	03 49 46.50	&	+05 51 42.3 	\phantom{aa}	&	BLRG	&	FRII	&	0.339	&	Feb-06	&	180.4	\phantom{aa}	&	545.3	\phantom{aa}	&	167.8	\phantom{aa}	\\
0349$-$27	&	\phantom{a}		&	 03 51 35.81	&	 -27 44 33.8 	\phantom{aa}	&	NLRG	&	FRII	&	0.066	&	Aug-05	&	180.4	\phantom{aa}	&	545.3	\phantom{aa}	&	167.8	\phantom{aa}	\\
0404$+$03	&	\phantom{a}	3C105	&	04 07 16.49	&	+03 42 25.8 	\phantom{aa}	&	NLRG	&	FRII	&	0.089	&	Sep-05	&	180.4	\phantom{aa}	&	545.3	\phantom{aa}	&	none	\phantom{aa}	\\
0409$-$75	&	\phantom{a}		&	04 08 48.49	&	 -75 07 19.3 	\phantom{aa}	&	NLRG	&	FRII	&	0.693	&	Aug-05	&	180.4	\phantom{aa}	&	545.3	\phantom{aa}	&	167.8	\phantom{aa}	\\
0442$-$28	&	\phantom{a}		&	 04 44 37.67	&	-28 09 54.6 	\phantom{aa}	&	NLRG	&	FRII	&	0.147	&	Feb-06	&	180.4	\phantom{aa}	&	545.3	\phantom{aa}	&	167.8	\phantom{aa}	\\
0521$-$36	&	\phantom{a}		&	05 22 58.94	&	-36 27 31.2 	\phantom{aa}	&	BLRG	&	C/J	&	0.055	&	Dec-05	&	92.3	\phantom{aa}	&	230.7	\phantom{aa}	&	167.8	\phantom{aa}	\\
0620$-$52	&	\phantom{a}		&	06 21 43.29 	&	 -52 41 33.3 	\phantom{aa}	&	WLRG	&	FRI	&	0.051	&	Dec-05	&	180.4	\phantom{aa}	&	545.3	\phantom{aa}	&	167.8	\phantom{aa}	\\
0625$-$35	&	\phantom{a}	OH-342	&	06 27 06.65 	&	 -35 29 16.3 	\phantom{aa}	&	WLRG	&	FRI	&	0.055	&	Dec-05	&	180.4	\phantom{aa}	&	545.3	\phantom{aa}	&	167.8	\phantom{aa}	\\
0625$-$53	&	\phantom{a}		&	06 26 20.44	&	-53 41 35.2 	\phantom{aa}	&	WLRG	&	FRII	&	0.054	&	Dec-05	&	180.4	\phantom{aa}	&	545.3	\phantom{aa}	&	167.8	\phantom{aa}	\\
0806$-$10	&	\phantom{a}	3C195	&	08 08 53.64	&	 -10 27 40.2 	\phantom{aa}	&	NLRG	&	FRII	&	0.110	&	Dec-05	&	180.4	\phantom{aa}	&	545.3	\phantom{aa}	&	167.8	\phantom{aa}	\\
0859$-$25	&	\phantom{a}		&	09 01 47.50	&	-25 55 19.0  	\phantom{aa}	&	NLRG	&	FRII	&	0.305	&	Dec-05	&	180.4	\phantom{aa}	&	545.3	\phantom{aa}	&	167.8	\phantom{aa}	\\
0915$-$11	&	\phantom{a}	Hydra A	&	09 18 05.67	&	-12 05 44.0 	\phantom{aa}	&	WLRG	&	FRI	&	0.054	&	May-04	&	165.7	\phantom{aa}	&	125.8	\phantom{aa}	&	83.9	\phantom{aa}	\\
0945$+$07	&	\phantom{a}	3C227	&	09 47 45.15	&	+07 25 20.4 	\phantom{aa}	&	BLRG	&	FRII	&	0.086	&	Dec-05	&	180.4	\phantom{aa}	&	545.3	\phantom{aa}	&	167.8	\phantom{aa}	\\
1136$-$13	&	\phantom{a}		&	11 39 10.75	&	-13 50 43.1 	\phantom{aa}	&	Q	&	FRII	&	0.554	&	Jun-06	&	180.4	\phantom{aa}	&	545.3	\phantom{aa}	&	167.8	\phantom{aa}	\\
1151$-$34	&	\phantom{a}		&	11 54 21.79	&	-35 05 29.1 	\phantom{aa}	&	Q	&	CSS	&	0.258	&	Feb-06	&	180.4	\phantom{aa}	&	543.3	\phantom{aa}	&	167.8	\phantom{aa}	\\
1226$+$02	&	\phantom{a}	3C273	&	12 29 06.70	&	+02 03 08.6 	\phantom{aa}	&	Q	&	C/J	&	0.158	&	Jun-04	&	48.2	\phantom{aa}	&	37.7	\phantom{aa}	&	41.9	\phantom{aa}	\\
1306$-$09	&	\phantom{a}		&	13 08 39.17	&	-09 50 32.6 	\phantom{aa}	&	NLRG	&	CSS	&	0.464	&	Feb-06	&	180.4	\phantom{aa}	&	545.3	\phantom{aa}	&	167.8	\phantom{aa}	\\
1355$-$41	&	\phantom{a}		&	13 59 00.23	&	  -41 52 54.1 	\phantom{aa}	&	Q	&	FRII	&	0.313	&	Feb-06	&	180.4	\phantom{aa}	&	545.3	\phantom{aa}	&	167.8	\phantom{aa}	\\
1547$-$79	&	\phantom{a}		&	 15 55 21.66	&	 -79 40 36.3  	\phantom{aa}	&	BLRG	&	FRII	&	0.483	&	Aug-05	&	180.4	\phantom{aa}	&	545.3	\phantom{aa}	&	167.8	\phantom{aa}	\\
1559$+$02	&	\phantom{a}	3C327	&	16 02 27.38	&	+01 57 55.7 	\phantom{aa}	&	NLRG	&	FRII	&	0.104	&	Mar-04	&	165.7	\phantom{aa}	&	125.8	\phantom{aa}	&	83.9	\phantom{aa}	\\
1602$+$01	&	\phantom{a}	3C327.1	&	16 04 45.35	&	+01 17 50.8 	\phantom{aa}	&	BLRG	&	FRII	&	0.462	&	Aug-05	&	180.4	\phantom{aa}	&	545.3	\phantom{aa}	&	167.8	\phantom{aa}	\\
1648$+$05	&	\phantom{a}	Herc A	&	16 51 08.16	&	+04 59 33.8 	\phantom{aa}	&	WLRG	&	FRI	&	0.154	&	Mar-04	&	165.7	\phantom{aa}	&	125.8	\phantom{aa}	&	83.9	\phantom{aa}	\\
1733$-$56	&	\phantom{a}		&	17 37 35.80	&	-56 34 03.4 	\phantom{aa}	&	BLRG	&	FRII	&	0.098	&	Aug-05	&	180.4	\phantom{aa}	&	545.3	\phantom{aa}	&	167.8	\phantom{aa}	\\
1814$-$63	&	\phantom{a}		&	18 19 34.96	&	 -63 45 48.1 	\phantom{aa}	&	NLRG	&	CSS	&	0.063	&	Aug-05	&	180.4	\phantom{aa}	&	545.3	\phantom{aa}	&	167.8	\phantom{aa}	\\
1839$-$48	&	\phantom{a}		&	18 43 14.64	&	-48 36 23.3 	\phantom{aa}	&	WLRG	&	FRI	&	0.112	&	Oct-06	&	180.4	\phantom{aa}	&	545.3	\phantom{aa}	&	none	\phantom{aa}	\\
1932$-$46	&	\phantom{a}		&	19 35 56.56	&	 -46 20 40.7 	\phantom{aa}	&	BLRG	&	FRII	&	0.231	&	May-06	&	180.4	\phantom{aa}	&	545.3	\phantom{aa}	&	167.8	\phantom{aa}	\\
1934$-$63	&	\phantom{a}		&	19 39 24.99	&	 -63 42 45.6 	\phantom{aa}	&	NLRG	&	GPS	&	0.183	&	May-06	&	180.4	\phantom{aa}	&	545.3	\phantom{aa}	&	167.8	\phantom{aa}	\\
1938$-$15	&	\phantom{a}		&	19 41 15.15	&	 -15 24 30.9 	\phantom{aa}	&	BLRG	&	FRII	&	0.452	&	Nov-06	&	180.4	\phantom{aa}	&	545.3	\phantom{aa}	&	none	\phantom{aa}	\\
1949$+$02	&	\phantom{a}	3C403	&	19 52 15.77	&	 +02 30 23.1 	\phantom{aa}	&	NLRG	&	FRII	&	0.059	&	Nov-05	&	92.3	\phantom{aa}	&	230.7	\phantom{aa}	&	167.8	\phantom{aa}	\\
1954$-$55	&	\phantom{a}		&	19 58 16.06	&	 -55 09 37.5 	\phantom{aa}	&	WLRG	&	FRI	&	0.060	&	May-06	&	180.4	\phantom{aa}	&	545.3	\phantom{aa}	&	167.8	\phantom{aa}	\\
2135$-$14	&	\phantom{a}		&	21 37 45.18 	&	-14 32 55.5 	\phantom{aa}	&	Q	&	FRII	&	0.200	&	Nov-05	&	180.4	\phantom{aa}	&	545.3	\phantom{aa}	&	167.8	\phantom{aa}	\\
2135$-$20	&	\phantom{a}	OX-258	&	21 37 50.00	&	 -20 42 31.7 	\phantom{aa}	&	BLRG	&	CSS	&	0.635	&	Nov-05	&	180.4	\phantom{aa}	&	545.3	\phantom{aa}	&	167.8	\phantom{aa}	\\
2211$-$17	&	\phantom{a}	3C444	&	22 14 25.76	&	 -17 01 36.2 	\phantom{aa}	&	WLRG	&	FRII	&	0.153	&	Dec-06	&	180.4	\phantom{aa}	&	545.3	\phantom{aa}	&	none	\phantom{aa}	\\
2221$-$02	&	\phantom{a}	3C445	&	22 23 49.57	&	 -02 06 12.4 	\phantom{aa}	&	BLRG	&	FRII	&	0.057	&	Nov-04	&	92.3	\phantom{aa}	&	69.2	\phantom{aa}	&	none	\phantom{aa}	\\
2250$-$41	&	\phantom{a}		&	22 53 03.16	&	 -40 57 46.2 	\phantom{aa}	&	NLRG	&	FRII	&	0.310	&	Nov-05	&	180.4	\phantom{aa}	&	545.3	\phantom{aa}	&	167.8	\phantom{aa}	\\
2314$+$03	&	\phantom{a}	3C459	&	 23 16 35.21	&	 +04 05 18.2 	\phantom{aa}	&	NLRG	&	FRII	&	0.220	&	Nov-05	&	92.3	\phantom{aa}	&	230.7	\phantom{aa}	&	167.8	\phantom{aa}	\\
2356$-$61	&	\phantom{a}		&	23 59 04.50	&	-60 54 59.1 	\phantom{aa}	&	NLRG	&	FRII	&	0.096	&	Dec-05	&	180.4	\phantom{aa}	&	545.3	\phantom{aa}	&	167.8	\phantom{aa}	\\

\enddata

\tablecomments{Table\ref{tbl-1} Column 5: broad line radio galaxy (BLRG), narrow line radio galaxy (NLRG), quasar (Q), weak line radio galaxy (WLRG, also known as low excitation galaxies). Column 6: Fanaroff-Riley class 1 (FRI), Fanaroff-Riley class 2 (FRII), compact steep spectrum type (CSS), core/jet (C/J), gigahertz peaked spectrum (GPS). Uncertain classification in brackets.}

\end{deluxetable}

\begin{deluxetable}{c@{\hspace{0mm}}l@{\hspace{0mm}}c@{\hspace{0mm}}r@{\hspace{0mm}}c@{\hspace{0mm}}r@{\hspace{0mm}}c@{\hspace{0mm}}r@{\hspace{0mm}}r}
\tabletypesize{\scriptsize}
\tablecaption{{\it Spitzer} Data Results \label{tbl-2}}
\tablewidth{0pt}
\tablehead{
\colhead{PKS Name}{\hspace{0mm}} & \colhead{Other name}{\hspace{0mm}} & \colhead{z}{\hspace{0mm}} & \colhead{$S_{24}$ (mJy)}{\hspace{0mm}} & \colhead{$\sigma$}{\hspace{0mm}} &
\colhead{$S_{70}$ (mJy)}{\hspace{0mm}} & \colhead{$\sigma$}{\hspace{0mm}} & \colhead{$S_{160}$ (mJy)}{\hspace{0mm}} & \colhead{$\sigma$} 
}
\startdata
0023$-$26	&	\phantom{aa}		&	0.322	&	2.4	\phantom{aaaaa}	&	0.3	&	26.3	\phantom{aaaaa}	&	3.1	&		83.6	\phantom{aaaaa}	&	16.7	\\
0034$-$01	&	\phantom{aa}	3C015	&	0.073	&	7.5	\phantom{aaaaa}	&	0.2	&	17.9	\phantom{aaaaa}	&	2.1	&	$<$	43.1	\phantom{aaaaa}	&	14.4	\\
0035$-$02	&	\phantom{aa}	3C17	&	0.220	&	12.2	\phantom{aaaaa}	&	0.1	&	23.6	\phantom{aaaaa}	&	4.8	&		97.2	\phantom{aaaaa}	&	9.1	\\
0038$+$09	&	\phantom{aa}	3C18	&	0.188	&	25.9	\phantom{aaaaa}	&	0.3	&	32.2	\phantom{aaaaa}	&	4.8	&	$<$	37.8	\phantom{aaaaa}	&	12.6	\\
0039$-$44	&	\phantom{aa}		&	0.346	&	33.0	\phantom{aaaaa}	&	0.4	&	68.7	\phantom{aaaaa}	&	4.6	&		84.6	\phantom{aaaaa}	&	13.5	\\
0043$-$42	&	\phantom{aa}		&	0.116	&	11.1	\phantom{aaaaa}	&	0.2	&	9.9	\phantom{aaaaa}	&	3.0	&	$<$	24.9	\phantom{aaaaa}	&	8.3	\\
0105$-$16	&	\phantom{aa}	3C32	&	0.400	&	9.7	\phantom{aaaaa}	&	0.2	&$<$	11.8	\phantom{aaaaa}	&	3.9	&	$<$	22.5	\phantom{aaaaa}	&	7.5	\\
0117$-$15	&	\phantom{aa}	3C38	&	0.565	&	6.1	\phantom{aaaaa}	&	0.2	&	20.2	\phantom{aaaaa}	&	1.8	&	$<$	46.4	\phantom{aaaaa}	&	15.5	\\
0213$-$13	&	\phantom{aa}	3C62	&	0.147	&	40.2	\phantom{aaaaa}	&	0.1	&	37.1	\phantom{aaaaa}	&	2.9	&	$<$	51.1	\phantom{aaaaa}	&	17.0	\\
0235$-$19	&	\phantom{aa}	OD-159	&	0.620	&	11.1	\phantom{aaaaa}	&	0.2	&	14.3	\phantom{aaaaa}	&	2.5	&	$<$	23.6	\phantom{aaaaa}	&	7.9	\\
0252$-$71	&	\phantom{aa}		&	0.566	&	2.9	\phantom{aaaaa}	&	0.1	&$<$	9.1	\phantom{aaaaa}	&	3.0	&	$<$	37.4	\phantom{aaaaa}	&	12.5	\\
0347$+$05	&	\phantom{aa}		&	0.339	&	3.5	\phantom{aaaaa}	&	0.2	&	30.8	\phantom{aaaaa}	&	4.3	&	$<$	52.2	\phantom{aaaaa}	&	17.4	\\
0349$-$27	&	\phantom{aa}		&	0.066	&	8.8	\phantom{aaaaa}	&	0.3	&	41.9	\phantom{aaaaa}	&	1.3	&	$<$	31.3	\phantom{aaaaa}	&	10.4	\\
0404$+$03	&	\phantom{aa}	3C105	&	0.089	&	30.8	\phantom{aaaaa}	&	0.1	&	70.9	\phantom{aaaaa}	&	2.0	&		-	\phantom{aaaaa}	&	-	\\
0409$-$75	&	\phantom{aa}		&	0.693	&	1.5	\phantom{aaaaa}	&	0.3	&	11.2	\phantom{aaaaa}	&	1.9	&	$<$	36.4	\phantom{aaaaa}	&	12.1	\\
0442$-$28	&	\phantom{aa}		&	0.147	&	22.0	\phantom{aaaaa}	&	0.3	&	31.0	\phantom{aaaaa}	&	4.5	&	$<$	25.9	\phantom{aaaaa}	&	8.6	\\
0521$-$36	&	\phantom{aa}		&	0.055	&	204.1	\phantom{aaaaa}	&	0.3	&	526.4	\phantom{aaaaa}	&	6.0	&		950.1	\phantom{aaaaa}	&	13.3	\\
0620$-$52	&	\phantom{aa}		&	0.051	&	4.5	\phantom{aaaaa}	&	0.1	&	47.3	\phantom{aaaaa}	&	1.4	&	$<$	24.3	\phantom{aaaaa}	&	8.1	\\
0625$-$35	&	\phantom{aa}	OH-342	&	0.055	&	24.7	\phantom{aaaaa}	&	0.3	&	44.8	\phantom{aaaaa}	&	1.3	&		131.0	\phantom{aaaaa}	&	11.2	\\
0625$-$53	&	\phantom{aa}		&	0.054	&	1.7	\phantom{aaaaa}	&	0.2	&$<$	10.8	\phantom{aaaaa}	&	3.6	&	$<$	52.6	\phantom{aaaaa}	&	17.5	\\
0806$-$10	&	\phantom{aa}	3C195	&	0.110	&	258.3	\phantom{aaaaa}	&	0.4	&	489.7	\phantom{aaaaa}	&	1.3	&		308.7	\phantom{aaaaa}	&	7.1	\\
0859$-$25	&	\phantom{aa}		&	0.305	&	9.3	\phantom{aaaaa}	&	0.4	&	8.4	\phantom{aaaaa}	&	2.8	&	$<$	52.3	\phantom{aaaaa}	&	17.4	\\
0915$-$11	&	\phantom{aa}	Hydra A	&	0.054	&	8.9	\phantom{aaaaa}	&	0.2	&	115.3	\phantom{aaaaa}	&	4.8	&		164.4	\phantom{aaaaa}	&	7.3	\\
0945$+$07	&	\phantom{aa}	3C227	&	0.086	&	47.7	\phantom{aaaaa}	&	0.3	&	19.4	\phantom{aaaaa}	&	3.5	&	$<$	44.1	\phantom{aaaaa}	&	14.7	\\
1136$-$13	&	\phantom{aa}		&	0.554	&	13.8	\phantom{aaaaa}	&	0.2	&	23.9	\phantom{aaaaa}	&	2.8	&	$<$	23.2	\phantom{aaaaa}	&	7.7	\\
1151$-$34	&	\phantom{aa}		&	0.258	&	16.4	\phantom{aaaaa}	&	0.3	&	51.5	\phantom{aaaaa}	&	3.0	&	$<$	46.3	\phantom{aaaaa}	&	15.4	\\
1226$+$02	&	\phantom{aa}	3C273	&	0.158	&	499.1	\phantom{aaaaa}	&	0.8	&	414.9	\phantom{aaaaa}	&	3.9	&		402.1	\phantom{aaaaa}	&	11.9	\\
1306$-$09	&	\phantom{aa}		&	0.464	&	4.6	\phantom{aaaaa}	&	0.2	&	21.7	\phantom{aaaaa}	&	2.1	&	$<$	29.2	\phantom{aaaaa}	&	9.7	\\
1355$-$41	&	\phantom{aa}		&	0.313	&	53.1	\phantom{aaaaa}	&	0.3	&	66.3	\phantom{aaaaa}	&	2.0	&	$<$	55.9	\phantom{aaaaa}	&	18.6	\\
1547$-$79	&	\phantom{aa}		&	0.483	&	7.9	\phantom{aaaaa}	&	0.1	&	19.2	\phantom{aaaaa}	&	1.7	&	$<$	50.2	\phantom{aaaaa}	&	16.7	\\
1559$+$02	&	\phantom{aa}	3C327	&	0.104	&	242.0	\phantom{aaaaa}	&	0.3	&	470.0	\phantom{aaaaa}	&	3.7	&		262.8	\phantom{aaaaa}	&	21.4	\\
1602$+$01	&	\phantom{aa}	3C327.1	&	0.462	&	7.7	\phantom{aaaaa}	&	0.3	&	12.3	\phantom{aaaaa}	&	2.5	&	$<$	32.4	\phantom{aaaaa}	&	10.8	\\
1648$+$05	&	\phantom{aa}	Herc A	&	0.154	&	2.0	\phantom{aaaaa}	&	0.2	&$<$	18.8	\phantom{aaaaa}	&	6.3	&		134.3	\phantom{aaaaa}	&	15.8	\\
1733$-$56	&	\phantom{aa}		&	0.098	&	29.2	\phantom{aaaaa}	&	0.3	&	151.0	\phantom{aaaaa}	&	3.3	&		318.2	\phantom{aaaaa}	&	10.9	\\
1814$-$63	&	\phantom{aa}		&	0.063	&	60.6	\phantom{aaaaa}	&	0.4	&	142.1	\phantom{aaaaa}	&	2.1	&		202.1	\phantom{aaaaa}	&	13.1	\\
1839$-$48	&	\phantom{aa}		&	0.112	&	3.1	\phantom{aaaaa}	&	0.3	&	10.9	\phantom{aaaaa}	&	2.5	&		-	\phantom{aaaaa}	&	-	\\
1932$-$46	&	\phantom{aa}		&	0.231	&	2.5	\phantom{aaaaa}	&	0.1	&	17.6	\phantom{aaaaa}	&	1.6	&	$<$	36.5	\phantom{aaaaa}	&	12.2	\\
1934$-$63	&	\phantom{aa}		&	0.183	&	17.4	\phantom{aaaaa}	&	0.1	&	19.9	\phantom{aaaaa}	&	2.2	&	$<$	33.2	\phantom{aaaaa}	&	11.1	\\
1938$-$15	&	\phantom{aa}		&	0.452	&	6.8	\phantom{aaaaa}	&	0.3	&	19.7	\phantom{aaaaa}	&	4.2	&		-	\phantom{aaaaa}	&	-	\\
1949$+$02	&	\phantom{aa}	3C403	&	0.059	&	193.0	\phantom{aaaaa}	&	0.2	&	348.4	\phantom{aaaaa}	&	3.7	&		251.3	\phantom{aaaaa}	&	7.1	\\
1954$-$55	&	\phantom{aa}		&	0.060	&	2.7	\phantom{aaaaa}	&	0.2	&	8.8	\phantom{aaaaa}	&	2.9	&	$<$	34.1	\phantom{aaaaa}	&	11.4	\\
2135$-$14	&	\phantom{aa}		&	0.200	&	104.9	\phantom{aaaaa}	&	0.2	&	113.7	\phantom{aaaaa}	&	4.8	&		118.6	\phantom{aaaaa}	&	20.5	\\
2135$-$20	&	\phantom{aa}	OX-258	&	0.635	&	4.3	\phantom{aaaaa}	&	0.3	&	37.0	\phantom{aaaaa}	&	4.4	&		127.1	\phantom{aaaaa}	&	10.3	\\
2211$-$17	&	\phantom{aa}	3C444	&	0.153	&	0.5	\phantom{aaaaa}	&	0.1	&$<$	9.4	\phantom{aaaaa}	&	3.1	&		-	\phantom{aaaaa}	&	-	\\
2221$-$02	&	\phantom{aa}	3C445	&	0.057	&	232.1	\phantom{aaaaa}	&	0.3	&	186.4	\phantom{aaaaa}	&	5.2	&		-	\phantom{aaaaa}	&	-	\\
2250$-$41	&	\phantom{aa}		&	0.310	&	11.6	\phantom{aaaaa}	&	0.1	&	22.0	\phantom{aaaaa}	&	2.5	&	$<$	27.6	\phantom{aaaaa}	&	9.2	\\
2314$+$03	&	\phantom{aa}	3C459	&	0.220	&	49.9	\phantom{aaaaa}	&	0.3	&	511.5	\phantom{aaaaa}	&	3.9	&		490.4	\phantom{aaaaa}	&	19.1	\\
2356$-$61	&	\phantom{aa}		&	0.096	&	41.0	\phantom{aaaaa}	&	0.2	&	74.7	\phantom{aaaaa}	&	2.2	&	$<$	35.4	\phantom{aaaaa}	&	11.8	\\

\enddata

\tablecomments{Table\ref{tbl-2}: All images reduced were from the BCD files, {\it Spitzer} Science Center pipeline version 14.4.0, Fluxes presented for 2135-20 are the mean fluxes of two identical observations see \S \ref{sec:MFIRnotes}.}

\end{deluxetable}

\subsection{24\moo\ Reduction}
As the shortest wavelength band, the 24\moo\ channel of {\it MIPS} is also the most
sensitive, and has the highest spatial resolution (6$''$ FWHM). Integration times for our targets were in the range 48.2
to 180.4 seconds, where over 80\% have the longer exposure time (see Table \ref{tbl-1}).

Minimal processing to the 24\moo\ SSC pipeline product is required
for most general science objectives. However, we re-processed the data
by mosaiking the BCD files using the MOPEX software and adding an
additional flatfield and overlap correction. The flatfield script is
part of the MOPEX software package. It computes a flat field from the
median of all the dithered frames, normalizes this to an average of 1
and divides into all the input BCD files. In contrast, the overlap
correction interpolates the input images onto a common grid and then
the cumulative pixel-by-pixel difference between the overlapping areas
is minimized. The final mosaic pixel size was set to 2.45 arsecs. This
was deliberately chosen to match the pipeline post-BCD products for
purposes of comparison. Examples of the images can be seen in Figure
\ref{1814}.

The 24\moo\ fluxes were extracted using aperture photometry in the
GAIA package. Aperture corrections were derived from an empirically
determined average curve of growth of flux vs. aperture diameter
derived from observations of bright sources in our sample with
$S_{24}>50$mJy (\object{PKS0521-36}, \object{PKS1226$+$02},
\object{PKS1949$+$02}, \object{PKS2221-02},
\object{PKS2314$+$03}). These aperture corrections are broadly
consistent with those published by the SSC. For most objects the
aperture was set to a standard radius of 15$''$ corresponding to an
aperture correction of 1.08. However, for a few objects an aperture of
half this size (7.5$''$) was used to avoid contamination from the flux
of objects close to the source. In the latter cases the applied
aperture correction was 1.52.

The standard deviation of 6 aperture flux measurements of background
sky patches, obtained using an identical aperture to the object flux
measurements, was used to derive the 1$\sigma$ uncertainties presented
in Table \ref{tbl-2}. These 1$\sigma$ values represent the
fluctuations in the sky background due to photon counting noise, as
well as mosaiking and flat fielding artifacts. Given that most of the
sources are faint relative to the background, these 1$\sigma$
measurements give a realistic estimate of the 1$\sigma$ uncertainties
in the flux measurement for the fainter sources in the sample. In
addition there is an estimated $\pm$4\% flux calibration uncertainty
\citep{engelbracht07}, which is likely to dominate for the brightest
sources in the sample. Overall, at 24\moo\ we detect 100\% of our
sample at the $>5\sigma$ level.

\subsection{70\moo\ Reduction} 
The 70\moo\ waveband is less sensitive than the 24\moo\ band and uses
an entirely different Ge:Ga detector technology, which is not as
stable as the Si:As detectors used at 24\moo. The spatial resolution
of {\it Spitzer}/{\it MIPS} at 70\moo\ is 18$''$(FWHM). Integration times for our targets
were in the range 37.7 to 545.3 seconds, where over 80\% have the longer exposure time (see Table
\ref{tbl-1}). The 70\moo\ data were also reduced with the MOPEX
package. In addition to the basic mosaiking process we used a column
filtering process from the SSC contributed software pages (IDL program
BCD column filter.pro), which calculates the median value for each
column and subtracts that from each column for each BCD. This process
significantly improves artifacts in the images with minimal loss of
flux from the source. The final mosaic pixel size is 3.9$''$, again
chosen to match that of the pipeline mosaic products.

The fluxes at 70\moo\ were also measured using aperture photometry,
with an aperture radius of 24$''$ for the majority of sources. Aperture
corrections and 1$\sigma$ uncertainties were derived in the same
manner as for the 24\moo\ images\footnote{N.B. In the case of 70\moo\
and 160\moo, in addition to the photon counting and instrumental
induced fluctuations, there will be an additional contribution to the
fluctuations in the background from sources close to the detection
limit. This will be reflected in the 1$\sigma$ estimates tabulated in
Table \ref{tbl-2}.}. Again these aperture corrections are broadly
consistent with those published by the SSC. In this case the aperture
correction factor is 1.33. A smaller aperture of half the size (22$''$)
was used for fainter sources to obtain better S/N for detections, with
a correction factor of 2.04. There is also an additional $\pm$10 \%
calibration error at 70\moo\ \citep{gordon07}. For undetected sources
upper limits were derived using the standard deviation of background
measurements to obtain $\sigma$. For these sources the values
presented in Table \ref{tbl-2} are the 3$\sigma$ upper limits, of
which there are 5.

Overall, we detect 90\% of our sample at 70\moo\ at the $>3\sigma$
level. 

\subsection{160\moo\ Reduction}
The longest wavelength {\it MIPS} band (160\moo) has the lowest effective
sensitivity and utilizes the same detector technology as 70\moo. The
spatial resolution of {\it Spitzer}/{\it MIPS} at 160\moo\ is 40$''$(FWHM). Integration
times for our targets at 160\moo\ were in the range 41.9 to 167.8
seconds, where over 90\% have the longer exposure time (see Table \ref{tbl-1}).

In this case we mosaiked the images with the MOPEX software (final
image pixel size 7.9$''$) with no additional processing, as none of
the available tools appeared to provide any significant improvement to
the images.

Aperture photometry proved impossible to use for extracting the
160\moo\ flux, due to the difficulty in estimating the background
accurately. This is because the PSF at 160\moo\ is on the scale of the
image, and there is also the strong possibility of contamination by
other sources in the field. Therefore, we used the APEX program in the
MOPEX GUI software to extract the fluxes of detected sources using PSF
fitting. The standard PSF available with the software was used for
this task. Overall we detect 33\% of our sample at 160\moo\ at the
$>3\sigma$ level.

The 1$\sigma$ errors were derived in an identical way to the other 2
wavebands, and the upper limits were derived in an identical way to
70\moo\ . There is also an additional $\pm$20\% flux calibration
uncertainty as stated in the {\it MIPS} data handbook (available from the
SSC website).  Upper limits were derived using the standard deviation
of background measurements to obtain $\sigma$ The values presented in
Table \ref{tbl-2} are the 3$\sigma$ upper limits, of which there are
27. Note that a few objects undetected by the APEX software appear to
be detected in a visual inspection of the image. Also, overlapping PSF
from nearby objects has prevented the APEX detection of a point source
in at least one image. Therefore, we believe that the APEX software
underestimates the true detection rate of sources in our sample; in
this sense the detection rate is conservative. This conclusion is
consistent with the fact that the typical 3$\sigma$ detection limit
for the 160\moo\ data is 55mJy, which is somewhat lower than the lowest
measured flux of 84.6mJy.

\subsection{Notes on the MFIR observations}
\label{sec:MFIRnotes}
Due to an error in the acquisition file for the data, PKS2135-20 was
observed twice. This gave us an opportunity to investigate the
uncertainty derived from two identical observations of the same
source. We found that the fluxes measured for this source in the two
data sets were consistent within 10\% in all {\it MIPS} bands. Fluxes
presented in Table 2 are the mean fluxes of the two observations.

It has also been possible to compare some of our flux values obtained
using {\it Spitzer} with those based on {\it IRAS} observations. Taking into account
BLRG/quasar objects that are potentially variable, we directly
compared the {\it IRAS} 25\moo\ and {\it Spitzer} 24\moo\ flux
measurement. However in order to make the comparison for the FIR
fluxes it was necessary to extrapolate the {\it IRAS} 60\moo\ fluxes to the
{\it Spitzer} 70\moo\ band using the {\it IRAS} 25 to 60\moo\ spectral index. We
find that the {\it Spitzer} and {\it IRAS} fluxes agree to within better than 20\%
at 24\moo\ and within better than 30\% at 70\moo. It is also worth
noting that our {\it Spitzer} measured flux values for sources in common
agree with published {\it Spitzer} fluxes in \citet{shi05} to within a few percent.

\section{Radio Observations and Reduction}

\label{sec:radio}

\begin{deluxetable}{@{\hspace{0mm}}c@{\hspace{0mm}}r@{\hspace{0mm}}c@{\hspace{0mm}}l@{\hspace{0mm}}l@{\hspace{0mm}}l@{\hspace{0mm}}l@{\hspace{0mm}}l@{\hspace{0mm}}l@{\hspace{0mm}}c@{\hspace{0mm}}r}
\tabletypesize{\scriptsize}
\tablecaption{{\it ATCA} Radio Core Data\label{tbl-3}}
\tablewidth{0pt}
\tablehead{
\colhead{PKS Name}{\hspace{0mm}} & \colhead{Radio Class}{\hspace{0mm}} & \colhead{z}{\hspace{0mm}} & \colhead{$S_{core}^{18Ghz}$(mJy)}{\hspace{0mm}} &\colhead{$\sigma$}{\hspace{0mm}} &\colhead{Beam($''$)}{\hspace{0mm}} & \colhead{$S_{core}^{24Ghz}$(mJy)}{\hspace{0mm}} & \colhead{$\sigma$}{\hspace{0mm}} &\colhead{Beam($''$)}{\hspace{0mm}}& \colhead{Combined}{\hspace{0mm}}& \colhead{R} 
}

\startdata
0039$-$44	&	FRII	\phantom{aaaa}	&	0.346	&	\phantom{aaa}	$<1.1$	&	0.4	&	\phantom{a}	0.5x0.3(-2.7)	&	\phantom{aaa}	$<0.9$	&	0.3	&	0.4x0.2(-2.8)	&	-	&	$<$0.0059	\\
0043$-$42	&	FRII	\phantom{aaaa}	&	0.116	&	\phantom{aaa}	-	&	0.3	&	\phantom{a}	0.6x0.3(26.4)	&	\phantom{aaa}	-	&	0.3	&	0.5x0.2(31.5)	&	1.4	&	0.0005	\\
0252$-$71	&	CSS	\phantom{aaaa}	&	0.566	&	\phantom{aaa}	278.1*	&	1.0	&	\phantom{a}	0.4x0.3(18.2)	&	\phantom{aaa}	187.7*	&	1.0	&	0.3x0.2(-20.8)	&	-	&	$<$0.0260	\\
0349$-$27	&	FRII	\phantom{aaaa}	&	0.066	&	\phantom{aaa}	20.0	&	0.1	&	\phantom{a}	0.7x0.3(3.8)	&	\phantom{aaa}	21.7	&	0.3	&	0.6x0.2(4.7)	&	-	&	0.0095	\\
0409$-$75	&	FRII	\phantom{aaaa}	&	0.693	&	\phantom{aaa}	$<10.0$	&	3.3	&	\phantom{a}	0.4x0.3(24.8)	&	\phantom{aaa}	$<9.0$	&	3.0	&	0.3x0.2(21.6)	&	-	&	$<$0.0023	\\
0442$-$28	&	FRII	\phantom{aaaa}	&	0.147	&	\phantom{aaa}	19.1	&	0.4	&	\phantom{a}	0.9x0.3(19.2)	&	\phantom{aaa}	20.4	&	0.4	&	0.6x0.2(17.4)	&	-	&	0.0090	\\
0521$-$36	&	C/J	\phantom{aaaa}	&	0.055	&	\phantom{aaa}	2798*	&	24.4	&	\phantom{a}	0.6x0.4(11.5)	&	\phantom{aaa}	2654*	&	14.8	&	0.4x0.3(14.5)	&	-	&	0.4115	\\
0620$-$52	&	FRI	\phantom{aaaa}	&	0.051	&	\phantom{aaa}	118.6	&	0.6	&	\phantom{a}	0.5x0.3(26.3)	&	\phantom{aaa}	101.2	&	0.7	&	0.4x0.2(28.6)	&	-	&	0.0964	\\
0625$-$35	&	FRI	\phantom{aaaa}	&	0.055	&	\phantom{aaa}	571.6	&	3.4	&	\phantom{a}	0.7x0.3(6.2)	&	\phantom{aaa}	478.8	&	3.0	&	0.5x0.2(7.5)	&	-	&	0.3293	\\
0625$-$53	&	FRII	\phantom{aaaa}	&	0.054	&	\phantom{aaa}	20.4	&	0.2	&	\phantom{a}	0.5x0.2(7.0)	&	\phantom{aaa}	19.3	&	0.2	&	0.3x0.2(8.6)	&	-	&	0.0130	\\
0859$-$25	&	FRII	\phantom{aaaa}	&	0.305	&	\phantom{aaa}	-	&	0.5	&	\phantom{a}	1.5x0.3(2.2)	&	\phantom{aaa}	-	&	0.5	&	1.1x0.2(3.6)	&	2.5	&	$<$0.0013	\\
1151$-$34	&	CSS	\phantom{aaaa}	&	0.258	&	\phantom{aaa}	759.9*	&	6.5	&	\phantom{a}	1.0x0.3(-23.4)	&	\phantom{aaa}	598.2*	&	6.3	&	0.7x0.2(-22.0)	&	-	&	$<$0.0072	\\
1355$-$41	&	FRII	\phantom{aaaa}	&	0.313	&	\phantom{aaa}	102.7	&	0.8	&	\phantom{a}	0.5x0.3(0.2)	&	\phantom{aaa}	89.6	&	1.0	&	0.4x0.3(-0.8)	&	-	&	0.0715	\\
1547$-$79	&	FRII	\phantom{aaaa}	&	0.483	&	\phantom{aaa}	-	&	0.2	&	\phantom{a}	0.6x0.3(-18.4)	&	\phantom{aaa}	-	&	0.2	&	0.5x0.2(-14.8)	&	1.4	&	0.0010	\\
1733$-$56	&	FRII	\phantom{aaaa}	&	0.098	&	\phantom{aaa}	293.6	&	0.9	&	\phantom{a}	0.4x0.3(-34.1)	&	\phantom{aaa}	269.4	&	1.9	&	0.3x0.2(-32.5)	&	-	&	0.0911	\\
1814$-$63	&	CSS	\phantom{aaaa}	&	0.063	&	\phantom{aaa}	1909*	&	6.5	&	\phantom{a}	0.4x0.3(-5.3)	&	\phantom{aaa}	1626*	&	7.3	&	0.3x0.2(-5.0)	&	-	&	0.0265	\\
1839$-$48	&	FRI	\phantom{aaaa}	&	0.112	&	\phantom{aaa}	108.2	&	0.5	&	\phantom{a}	0.5x0.3(-3.3)	&	\phantom{aaa}	100.9	&	0.6	&	0.4x0.2(-3.3)	&	-	&	0.0889	\\
1932$-$46	&	FRII	\phantom{aaaa}	&	0.231	&	\phantom{aaa}	16.4	&	0.7	&	\phantom{a}	0.5x0.4(-67.4)	&	\phantom{aaa}	21.5	&	0.5	&	0.4x0.3(-65)	&	-	&	0.0050	\\
1934$-$63	&	GPS	\phantom{aaaa}	&	0.183	&	\phantom{aaa}	1045*	&	2.0	&	\phantom{a}	0.6x0.3(68.9)	&	\phantom{aaa}	755.1*	&	1.1	&	0.6x0.2(66.3)	&	-	&	-	\\
1954$-$55	&	FRI	\phantom{aaaa}	&	0.060	&	\phantom{aaa}	61.7	&	0.2	&	\phantom{a}	0.4x0.3(56.6)	&	\phantom{aaa}	61.6	&	0.3	&	0.3x0.3(63.8)	&	-	&	0.0393	\\
2250$-$41	&	FRII	\phantom{aaaa}	&	0.310	&	\phantom{aaa}	$<1.9$	&	0.4	&	\phantom{a}	0.7x0.3(-20.2)	&	\phantom{aaa}	$<1.1$	&	0.4	&	0.5x0.3(-19.2)	&	-	&	$<$0.0012	\\
2356$-$61	&	FRII	\phantom{aaaa}	&	0.096	&	\phantom{aaa}	59.7	&	0.3	&	\phantom{a}	0.4x0.3(76.6)	&	\phantom{aaa}	63.8	&	0.4	&	0.3x0.2(89.4)	&	-	&	0.0137	\\

\enddata

\tablecomments{Table\ref{tbl-3}: Flux density results from the {\it ATCA} observations. Definitions for column 2 see Table\ref{tbl-1}. (*) denotes a total flux measurement as the resolution of the observations is not high enough to resolve the core in these compact objects. Column 11 gives the R parameter defined as $S_{core}/(S_{tot}-S_{core})$ where the total flux is taken at 5GHz assuming a flat spectrum.}

\end{deluxetable}

\begin{deluxetable}{c@{\hspace{-3mm}}l@{\hspace{-3mm}}c@{\hspace{-3mm}}c@{\hspace{-2mm}}r@{\hspace{1mm}}c@{\hspace{1mm}}l@{\hspace{-2mm}}r@{\hspace{1mm}}c@{\hspace{1mm}}l@{\hspace{0mm}}r}
\tabletypesize{\scriptsize}
\tablecaption{{\it VLA} Radio Core Data\label{tbl-4}}
\tablewidth{0pt}
\tablehead{
\colhead{PKS Name}{\hspace{-3mm}} & \colhead{Other Name}{\hspace{-3mm}} & \colhead{Radio Class}{\hspace{-3mm}} & \colhead{z}{\hspace{-2mm}} & \colhead{$S_{core}^{14.9Ghz}$(mJy)}{\hspace{-1mm}}&\colhead{$\sigma$}{\hspace{-1mm}} &
\colhead{Beam($''$)}{\hspace{-2mm}} & \colhead{$S_{core}^{?Ghz}$(mJy)}{\hspace{-1mm}}&\colhead{$\sigma$}{\hspace{-1mm}} & \colhead{Beam($''$)}{\hspace{-1mm}}& \colhead{R} 
}
\startdata
0023$-$26	&	\phantom{aa}		&	CSS	&	0.322	&	995.8*	\phantom{aaaaa}	&	7.0	&	\phantom{a}	0.7x0.3(-7.1)	&	820*	\phantom{aaaa}	&	9.0	&	\phantom{a}	0.5x0.2(-10.2)	&	$<$0.0059	\\
0034$-$01	&	\phantom{aa}	3C015	&	FRII	&	0.073	&	31.5	\phantom{aaaaa}	&	0.5	&	\phantom{a}	0.4x0.3(51.4)	&	36.9	\phantom{aaaa}	&	0.9	&	\phantom{a}	0.3x0.2(53.7)	&	0.0218	\\
0035$-$02	&	\phantom{aa}	3C17	&	(FRII)	&	0.220	&	320.6	\phantom{aaaaa}	&	1.9	&	\phantom{a}	0.4x0.4(45.7)	&	249.8	\phantom{aaaa}	&	4.2	&	\phantom{a}	0.2x0.2(-79.2)	&	0.2870	\\
0038$+$09	&	\phantom{aa}	3C18	&	FRII	&	0.188	&	55.6	\phantom{aaaaa}	&	0.5	&	\phantom{a}	0.3x0.3(-51.1)	&	68.2	\phantom{aaaa}	&	0.9	&	\phantom{a}	0.2x0.2(51.4)	&	0.0387	\\
0105$-$16	&	\phantom{aa}	3C32	&	FRII	&	0.400	&	$<1.1$	\phantom{aaaaa}	&	0.4	&	\phantom{a}	0.5x0.3(4.8)	&	$<2.3$	\phantom{aaaa}	&	0.8	&	\phantom{a}	0.3x0.2(5.9)	&	$<$0.0015	\\
0117$-$15	&	\phantom{aa}	3C38	&	FRII	&	0.565	&	$<2.4$	\phantom{aaaaa}	&	0.8	&	\phantom{a}	0.5x0.3(6.6)	&	$<2.8$	\phantom{aaaa}	&	0.9	&	\phantom{a}	0.3x0.2(10.3)	&	0.0019	\\
0235$-$19	&	\phantom{aa}	OD-159	&	FRII	&	0.620	&	$<1.7$	\phantom{aaaaa}	&	0.6	&	\phantom{a}	0.6x0.3(8.9)	&	$<2.8$	\phantom{aaaa}	&	0.9	&	\phantom{a}	0.4x0.1(11.1)	&	$<$0.0016	\\
0347$+$05	&	\phantom{aa}		&	FRII	&	0.339	&	$<2.2$	\phantom{aaaaa}	&	0.7	&	\phantom{a}	0.4x0.3(-55.9)	&	$<2.5$	\phantom{aaaa}	&	0.8	&	\phantom{a}	0.2x0.2(-62)	&	$<$0.0019	\\
0349$-$27	&	\phantom{aa}		&	FRII	&	0.066	&	17.5	\phantom{aaaaa}	&	0.4	&	\phantom{a}	0.8x0.3(15.4)	&	17.4	\phantom{aaaa}	&	1.1	&	\phantom{a}	0.6x0.2(15.7)	&	0.0095	\\
0404$+$03	&	\phantom{aa}	3C105	&	FRII	&	0.089	&	19.8	\phantom{aaaaa}	&	0.5	&	\phantom{a}	0.4x0.3(-25.5)	&	?	\phantom{aaaa}	&	-	&	\phantom{a}	0.3x0.2(63.9)	&	0.0083	\\
0806$-$10	&	\phantom{aa}	3C195	&	FRII	&	0.110	&	32.7	\phantom{aaaaa}	&	0.4	&	\phantom{a}	0.4x0.3(-1.6)	&	20.7	\phantom{aaaa}	&	0.4	&	\phantom{a}	0.3x0.2(-4.0)	&	0.0166	\\
0859$-$25	&	\phantom{aa}		&	FRII	&	0.305	&	$<1.9$	\phantom{aaaaa}	&	0.6	&	\phantom{a}	0.7x0.3(0.6)	&	$<2.5$	\phantom{aaaa}	&	0.8	&	\phantom{a}	0.5x0.2(2.3)	&	$<$0.0013	\\
0915$-$11	&	\phantom{aa}	Hydra A	&	FRI	&	0.054	&	211.3	\phantom{aaaaa}	&	2.5	&	\phantom{a}	0.5x0.3(-1.8)	&	184.5	\phantom{aaaa}	&	1.7	&	\phantom{a}	0.3x0.2(-2.0)	&	0.0146	\\
0945$+$07	&	\phantom{aa}	3C227	&	FRII	&	0.086	&	13.9	\phantom{aaaaa}	&	0.3	&	\phantom{a}	0.4x0.3(-32.2)	&	11.7	\phantom{aaaa}	&	0.3	&	\phantom{a}	0.3x0.1(-23.3)	&	0.0049	\\
1136$-$13	&	\phantom{aa}		&	FRII	&	0.554	&	234.7	\phantom{aaaaa}	&	1.4	&	\phantom{a}	0.5x0.3(-0.5)	&	230.4	\phantom{aaaa}	&	1.6	&	\phantom{a}	0.3x0.2(-1.8)	&	0.1395	\\
1226$+$02	&	\phantom{aa}	3C273	&	C/J	&	0.158	&	24060.0	\phantom{aaaaa}	&	102.0	&	\phantom{a}	0.4x0.3(-46.7)	&	19460	\phantom{aaaa}	&	185	&	\phantom{a}	0.2x0.2(-59.1)	&	1.1884	\\
1306$-$09	&	\phantom{aa}		&	CSS	&	0.464	&	505.4*	\phantom{aaaaa}	&	4.7	&	\phantom{a}	0.4x0.3(-2.5)	&	397.9*	\phantom{aaaa}	&	3.8	&	\phantom{a}	0.3x0.2(2.8)	&	$<$0.0152	\\
1559$+$02	&	\phantom{aa}	3C327	&	FRII	&	0.104	&	15.2	\phantom{aaaaa}	&	0.3	&	\phantom{a}	0.4x0.3(44.3)	&	11.2	\phantom{aaaa}	&	0.3	&	\phantom{a}	0.3x0.2(53.8)	&	0.0046	\\
1602$+$01	&	\phantom{aa}	3C327.1	&	FRII	&	0.462	&	63.7	\phantom{aaaaa}	&	0.6	&	\phantom{a}	0.4x0.3(50.9)	&	65.5	\phantom{aaaa}	&	0.7	&	\phantom{a}	0.3x0.2(58.3)	&	0.0734	\\
1648$+$05	&	\phantom{aa}	Herc A	&	FRI	&	0.154	&	$4.5^a$	\phantom{aaaaa}	&	-	&	\phantom{a}	-	&	-	\phantom{aaaa}	&	-	&	\phantom{a}	-	&	0.0004	\\
1949$+$02	&	\phantom{aa}	3C403	&	FRII	&	0.059	&	20.0	\phantom{aaaaa}	&	0.6	&	\phantom{a}	0.4x0.4(53.6)	&	8.5	\phantom{aaaa}	&	0.4	&	\phantom{a}	0.3x0.2(53.6)	&	0.0060	\\
2135$-$14	&	\phantom{aa}		&	FRII	&	0.200	&	106.3	\phantom{aaaaa}	&	1.1	&	\phantom{a}	0.5x0.3(-7.7)	&	44.3	\phantom{aaaa}	&	1.1	&	\phantom{a}	0.3x0.2(-14.1)	&	0.0577	\\
2135$-$20	&	\phantom{aa}	OX-258	&	CSS	&	0.635	&	416*	\phantom{aaaaa}	&	3.8	&	\phantom{a}	0.6x0.3(-3.1)	&	199.8*	\phantom{aaaa}	&	3.3	&	\phantom{a}	0.4x0.2(-7.1)	&	0.0076	\\
2211$-$17	&	\phantom{aa}	3C444	&	FRII	&	0.153	&	$<1.3$	\phantom{aaaaa}	&	0.4	&	\phantom{a}	0.5x0.3(1.1)	&	$<1.8$	\phantom{aaaa}	&	0.6	&	\phantom{a}	0.5x0.29(-7.6)	&	$<$0.0007	\\
2221$-$02	&	\phantom{aa}	3C445	&	FRII	&	0.057	&	27.3	\phantom{aaaaa}	&	0.4	&	\phantom{a}	0.4x0.3(46.6)	&	16.1	\phantom{aaaa}	&	0.5	&	\phantom{a}	0.3x0.2(14.5)	&	0.0096	\\
2314$+$03	&	\phantom{aa}	3C459	&	FRII	&	0.220	&	$<162.3^c$	\phantom{aaaaa}	&	1.0	&	\phantom{a}	0.4x0.3(49.9)	&	$<92.6^c$	\phantom{aaaa}	&	0.9	&	\phantom{a}	0.3x0.2(58.5)	&	$<$0.1060	\\
\cutinhead{VLBI core upper limits of compact objects in the sample from \cite{tzioumis02}}																										
0023$-$26	&	\phantom{aa}	CSS	&	$<20 mJy$	&	at 2.3GHz	&		\phantom{aaaaa}	&		&	\phantom{a}		&		\phantom{aaaa}	&		&	\phantom{a}		&		\\
0252$-$71	&	\phantom{aa}	CSS	&	$<40 mJy$	&	at 2.3GHz	&		\phantom{aaaaa}	&		&	\phantom{a}		&		\phantom{aaaa}	&		&	\phantom{a}		&		\\
1151$-$34	&	\phantom{aa}	CSS	&	$<20 mJy$	&	at 8.3GHz	&		\phantom{aaaaa}	&		&	\phantom{a}		&		\phantom{aaaa}	&		&	\phantom{a}		&		\\
1306$-$09	&	\phantom{aa}	CSS	&	$<28.5 mJy$	&	at 2.3GHz	&		\phantom{aaaaa}	&		&	\phantom{a}		&		\phantom{aaaa}	&		&	\phantom{a}		&		\\
1814$-$63	&	\phantom{aa}	CSS	&	$87 mJy^b$	&	at 2.3GHz	&		\phantom{aaaaa}	&		&	\phantom{a}		&		\phantom{aaaa}	&		&	\phantom{a}		&		\\
2135$-$20	&	\phantom{aa}	CSS	&	$<11.5 mJy$	&	at 5GHz	&		\phantom{aaaaa}	&		&	\phantom{a}		&		\phantom{aaaa}	&		&	\phantom{a}		&		\\

\enddata

\tablecomments{Table\ref{tbl-4}: Flux density results from {\it VLA} observations. (*) total flux measurement as the resolution of the observations is not high enough to resolve the core in these compact objects. }
\tablenotetext{a}{8.4Ghz flux taken from \citet{gizani03}}
\tablenotetext{b}{Possible detection at the 15$\sigma$ level. }
\tablenotetext{c}{VLBI radio maps of PKS2314+03 presented in \citet{tomasson03} resolve the central region of this object into two components of which it is unclear which is the true flat spectrum core. Our images cannot resolve this central structure and therefore the flux measurement presented here is likely to be an upper limit of the true flat spectrum core flux. }

\end{deluxetable}

\subsection{Sample}
 The radio sample is identical to that discussed in section
 \ref{sec:sample}. The objects were divided between the {\it ATCA} and the
 {\it VLA} observatories according to the declination of the source (cutoff
 $\delta\approx-25\degree$). Four objects -- \object{PKS0023-26},
 \object{PKS0349-27}, \object{PKS0442-28}, \object{PKS0859-25}-- were
 observed with both {\it ATCA} and the {\it VLA}.

\begin{figure*}
\epsscale{1.4}
\plotone{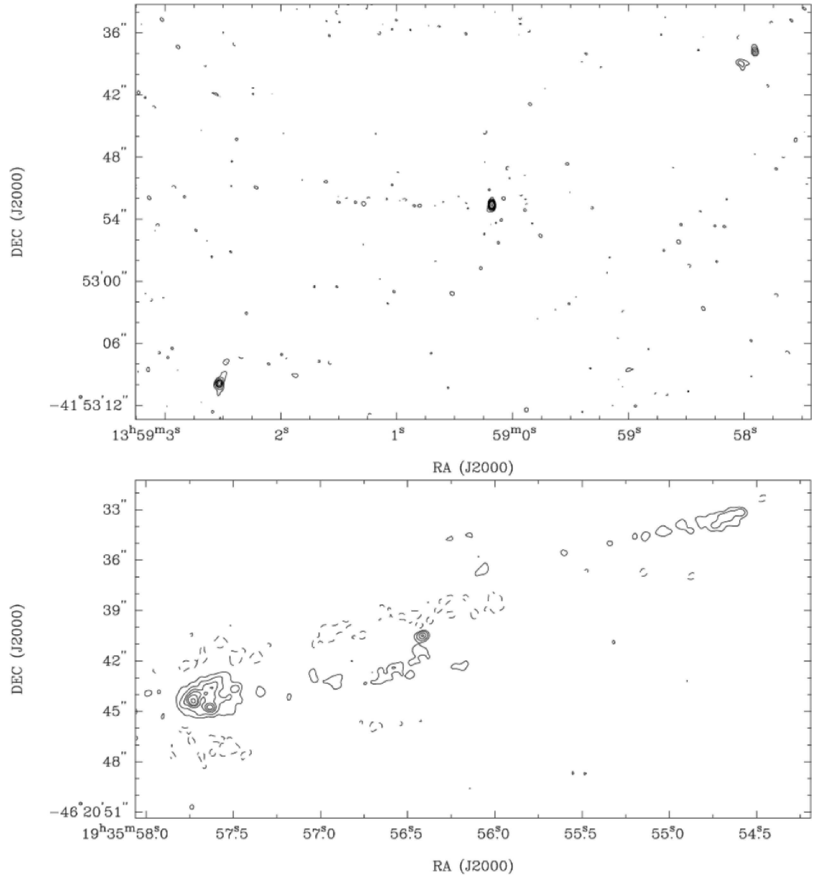}
\caption{Example images from our high frequency radio
observations. Top: PKS1355$-$41, an FRII quasar. Bottom: PKS1932$-$46,
an FRII broad line radio galaxy. Contour level for the plots are as
follows: 2.4*-1, 1, 2, 4, 6, 8, 12, 16, 32, 64, 128 mJy/beam. The core
and the lobes/hot spots are clearly seen in both objects. In PKS
1932-46, residuals are seen on large scales due to extended, low
surface brightness structures that cannot be properly imaged due to
the poor uv coverage. \label{fig:rad_image}}
\end{figure*}

\subsection{{\it ATCA} Observations}

We observed 23 objects (with $\delta < -25\degree$) from our sample
with {\it ATCA}. For one object (\object{PKS0023$-$26}) the quality of the {\it ATCA} data 
turned out to be quite low, we have therefore excluded it from our {\it ATCA} 
list and we will use only the {\it VLA} data.

The sources were observed using the 6-km configuration --
the longest available for this radio telescope.  These observations
were carried out between the 19th and 23rd of July 2006 in perfect
weather conditions. The total run comprised of a continuous 100h of observations.
 
Data were taken simultaneously at 17.9 GHz and 24.1 GHz.  These high
frequencies allow us to achieve a relatively high resolution.  Each
source was observed for a period ranging from 30 to 60 min (depending
on scheduling constraints) in scans of 10 min spread over a period of
12h in order to obtain a good enough {\sl uv}-coverage.  In each scan
we swapped between the two frequencies. Furthermore, for each
frequency, we actually took data simultaneously at 17.47 and 18.50 GHz
in the first observation and at 23.62 and 24.64 GHz in the second,
with a bandwidth of 128 MHz for each of these frequencies.  These
separate frequencies allowed us to improve (radially) the $uv$
coverage.  The flux calibrator -- PKS1934-64 flux level at 12mm =
1.03Jy -- was observed few times during the whole run.  In addition, a
nearby phase calibrator was observed for 5 min before every scan on a
source.  A reference pointing observation was also carried out before every source. 
 
The {\it ATCA} data were calibrated using the MIRIAD software package
\citep{sault95}.  The final images were obtained using multi-frequency
synthesis option and adding together the two nearby frequencies (17.47
and 18.50 GHz; 23.62 and 24.64 GHz) to increase the sensitivity and
improve the {\sl uv}-coverage.  Typically two cycles of
self-calibration were needed to obtain the final images.  In these
images, we have used uniform weighting to achieve the highest possible
resolution.

Table \ref{tbl-3} summarises the results of the {\it ATCA} observations,
including beam size and noise for each object and each frequency.  We
have detected radio cores in 14 of the 22 sources (63\%; PKS~0023-26
is excluded from the {\it ATCA} list as discussed above).  In a few cases,
in particular those for which the core was not detected, we have
combined both frequencies (17 and 24 GHz) in order to improve the
sensitivity (\object{PKS0043-42}, \object{PKS0859-25},
\object{PKS1547-79}).  Upper limits to the flux and power of the
undetected cores, were estimated using $3\sigma$ of the noise level.
A few objects are known to be compact steep spectrum
(\object{PKS0252-71}, \object{PKS1151-34}, \object{PKS1814-63},
\object{PKS1934-63}). These sources are already part of the list of
{\it ATCA} calibrators and indeed the resolution of our observation is not good
enough to pinpoint a possible core. We have used these objects as a
check of on our measured flux level (they agree very well with the data
from the calibration archive and they show no indication of
variability). For these sources, core fluxes or upper limit have been
measured from VLBI maps in \cite{tzioumis02} (listed at
the end of Table \ref{tbl-4}).

Finally, in a number of objects (PKS0039-44, PKS0043-42, PKS0409-75,
PKS0442-28, PKS0859-25, PKS1355-41, PKS1547-79, PKS1932-46, PKS
2250-41) we also detected radio emission from other regions of the
radio galaxy and, in particular, from the hot spots. Two examples of
this are shown in Fig. \ref{fig:rad_image}. These maps illustrate the
fact that at the high radio frequency and resolution of the
observations, the maps are dominated by the compact cores, hotspots
and knot features along the jets, and the diffuse lobe emission is
resolved out.

\subsection{{\it VLA} Observations}
We observed 27 objects with the {\it VLA} at frequencies of 14.9 and
22.4~GHz. The observations were carried out between 4 Aug, 2-3 Sep
2006 in B-array configuration.  Due to servicing, 10 out of 27 dishes
were not operating during the observation period. For one object
(\object{PKS0442$-$28}) the quality of the {\it VLA} data turned out to be
quite low due to the declination of the source, we therefore have
excluded it from our {\it VLA} list and we will use only the {\it ATCA} data.

Each source was observed for about 15 min per frequency, separated in
three scans of 5~min each. In addition, a reference pointing
observation was carried out before every source. The observations at
the two frequencies were interleaved, to provide a better {\sl
uv}-coverage.  As flux calibrators we used PKS0713$+$438 for the
U-band (14.4-14.9GHz), and PKS1331$+$305 (3C286) for the K-band
(22.1-26.0GHz). These calibrators were observed a few times during
the run.  Data splitting and opacity corrections were done inside
AIPS, while the rest of the data reduction was done using the MIRIAD
software package.  Images of the separate frequencies were obtained
for each source and combined images were also made in order to attempt
to detect those objects with faint cores, to no avail.  Table
\ref{tbl-4} summarises the results of the {\it VLA} observations.

Overall, from the {\it VLA} observations, we detected cores in 58\% of the
sources (15 cores detected in the 26 successfully observed).

Again, in 7 of the {\it VLA} objects (PKS0117-15, PKS0347-05,
PKS0859-25, PKS0915-11, PKS 1136$+$07, PKS1602-01, PKS2314-03)
we also detected radio emission from extended regions of the radio
galaxy and, in particular, from the hot spots.

\subsection{Notes on the radio observations}
We chose a high resolution array configuration in order to try and
detect the high frequency radio cores. Due to this the resolution of
the images resolves out the extended structure of the
sources. Therefore it would be impossible to provide accurate
measurements for the total emission from our sources with these
observations at 15 to 22 GHz.

Also, in the final column of the {\it ATCA} and {\it VLA} data tables we present
estimates of the orientation sensitive R parameter defined as
$S_{core}/(S_{tot}-S_{core})$. In order to determine this we have used
estimates of the total radio flux measured at 5GHz $S_{tot}$ taken from \citet{morganti93}, and the
core measurement is taken from the lower of the two observed frequency
core detections of either the {\it ATCA} (18 GHz) or {\it VLA} (15 GHz)
observations. Here we assume that the flat spectrum core flux remained
constant between 5 GHz and the higher frequencies. The analysis of
these data will be addressed in the detailed discussion paper to
follow.

\begin{deluxetable}{c@{\hspace{0mm}}c@{\hspace{-2mm}}c@{\hspace{-2mm}}c@{\hspace{-2mm}}c@{\hspace{-2mm}}c@{\hspace{1mm}}c@{\hspace{0mm}}c@{\hspace{-2mm}}c@{\hspace{-2mm}}c@{\hspace{-2mm}}c@{\hspace{-2mm}}c}
\tabletypesize{\scriptsize}
\tablecaption{SED Truth Table\label{tbl-5}}
\tablewidth{0pt}
\tablehead{
\colhead{Name}{\hspace{0mm}} & \colhead{Other}{\hspace{-2mm}} & \colhead{Opt.Class}{\hspace{-2mm}} & \colhead{Rad.Class}{\hspace{-2mm}} & \colhead{Total}{\hspace{-2mm}} &\colhead{Core}{\hspace{1mm}}& \colhead{Name}{\hspace{0mm}} & \colhead{Other}{\hspace{-2mm}} & \colhead{Opt.Class}{\hspace{-2mm}} & \colhead{Rad.Class}{\hspace{-2mm}} & \colhead{Total}{\hspace{-2mm}} &\colhead{Core}
}

\startdata
0023$-$26	&		&	NLRG	&	CSS	&	$X$	&	$X$	&	1151$-$34	&		&	Q	&	CSS	&	$X$	&	$X$	\\
0034$-$01	&	3C015	&	WLRG	&	FRII	&	$X$	&	\checkmark	&	1226$+$02	&	3C273	&	Q	&	C/J	&	-	&	-	\\
0035$-$02	&	3C17	&	BLRG	&	(FRII)	&	\checkmark	&	\checkmark	&	1306$-$09	&		&	NLRG	&	CSS	&	\checkmark	&	$X$	\\
0038$+$09	&	3C18	&	BLRG	&	FRII	&	$X$	&	\checkmark	&	1355$-$41	&		&	Q	&	FRII	&	$X$	&	$X$	\\
0039$-$44	&		&	NLRG	&	FRII	&	$X$	&	$X$	&	1547$-$79	&		&	BLRG	&	FRII	&	$X$	&	$X$	\\
0043$-$42	&		&	WLRG	&	FRII	&	$X$	&	$X$	&	1549$-$79	&		&	NLRG	&	CFS	&	-	&	-	\\
0105$-$16	&	3C32	&	NLRG	&	FRII	&	$X$	&	$X$	&	1559$+$02	&	3C327	&	NLRG	&	FRII	&	$X$	&	$X$	\\
0117$-$15	&	3C38	&	NLRG	&	FRII	&	$X$	&	$X$	&	1602$+$01	&	3C327.1	&	BLRG	&	FRII	&	$X$	&	\checkmark	\\
0213$-$13	&	3C62	&	NLRG	&	FRII	&	$X$	&	$X$	&	1648$+$05	&	Herc A	&	WLRG	&	FRI	&	$X$	&	$X$	\\
0235$-$19	&	OD-159	&	BLRG	&	FRII	&	$X$	&	$X$	&	1733$-$56	&		&	BLRG	&	FRII	&	$X$	&	$X$	\\
0252$-$71	&		&	NLRG	&	CSS	&	$X$	&	$X$	&	1814$-$63	&		&	NLRG	&	CSS	&	$X$	&	$X$	\\
0347$+$05	&		&	BLRG	&	FRII	&	$X$	&	$X$	&	1839$-$48	&		&	WLRG	&	FRI	&	\checkmark	&	\checkmark	\\
0349$-$27	&		&	NLRG	&	FRII	&	$X$	&	$X$	&	1932$-$46	&		&	BLRG	&	FRII	&	$X$	&	\checkmark	\\
0404$+$03	&	3C105	&	NLRG	&	FRII	&	$X$	&	$X$	&	1934$-$63	&		&	NLRG	&	GPS	&	$X$	&	$X$	\\
0409$-$75	&		&	NLRG	&	FRII	&	\checkmark	&	$X$	&	1938$-$15	&		&	BLRG	&	FRII	&	$X$	&	-	\\
0442$-$28	&		&	NLRG	&	FRII	&	$X$	&	$X$	&	1949$+$02	&	3C403	&	NLRG	&	FRII	&	$X$	&	$X$	\\
0521$-$36	&		&	BLRG	&	C/J	&	-	&	-	&	1954$-$55	&		&	WLRG	&	FRI	&	\checkmark	&	\checkmark	\\
0620$-$52	&		&	WLRG	&	FRI	&	$X$	&	\checkmark	&	2135$-$14	&		&	Q	&	FRII	&	$X$	&	$X$	\\
0625$-$35	&	OH-342	&	WLRG	&	FRI	&	$X$	&	\checkmark	&	2135$-$20	&	O$X$-258	&	BLRG	&	CSS	&	$X$	&	$X$	\\
0625$-$53	&		&	WLRG	&	FRII	&	$X$	&	\checkmark	&	2211$-$17	&	3C444	&	WLRG	&	FRII	&	$X$	&	$X$	\\
0806$-$10	&	3C195	&	NLRG	&	FRII	&	$X$	&	$X$	&	2221$-$02	&	3C445	&	BLRG	&	FRII	&	$X$	&	$X$	\\
0859$-$25	&		&	NLRG	&	FRII	&	$X$	&	$X$	&	2250$-$41	&		&	NLRG	&	FRII	&	$X$	&	$X$	\\
0915$-$11	&	Hydra A	&	WLRG	&	FRI	&	$X$	&	$X$	&	2314$+$03	&	3C459	&	NLRG	&	FRII	&	$X$	&	$X$	\\
0945$+$07	&	3C227	&	BLRG	&	FRII	&	$X$	&	$X$	&	2356$-$61	&		&	NLRG	&	FRII	&	$X$	&	$X$	\\
1136$-$13	&		&	Q	&	FRII	&	\checkmark	&	\checkmark	&		&		&		&		&		&		\\

\enddata

\tablecomments{Table\ref{tbl-5}: Truth table investigating the potential non-thermal contamination of the thermal MFIR emission, see \S \ref{sec:lobe} and \S \ref{sec:core}. }

\end{deluxetable}

\section{Non-thermal Contamination}
\label{sec:sed}
MFIR emission can potentially have a thermal and/or non-thermal
origin in radio loud AGN: thermal emission from re-radiation of AGN
light by dust, and non-thermal synchrotron radiation. Therefore, in
order to investigate the MFIR dust emission we need to be aware of the
potential contamination by non-thermal sources. There are two important
components to consider when investigating the non-thermal
contamination in a sample of steep-spectrum radio-loud AGN.

\begin{itemize}
\item{\bf Total steep spectrum emission:} Steep spectrum components
  such as lobes and hotspots could be potential non-thermal
  contaminants of the MFIR emission. Despite the fact that these components fall
  in flux towards higher frequencies, they could still potentially
  dominate the flux of an object at MFIR wavelengths. However, the
  proportion of this non-thermal contamination also depends on how
  much of the steep spectrum emission regions lies within the MFIR
  instrument beam (in this case the {\it Spitzer} beam). Thus, the degree of
  contamination by steep spectrum components for most of the nearby extended
  sources in our sample will be insignificant because they have a high
  proportion of their synchrotron emitting lobes/hotspots far outside
  the beam. The contamination is likely to be most important for
  compact objects such as CSS/GPS, and for extended FRII sources at
  high redshift that appear small because of their large
  distances. At the average redshift (z = 0.244) of our complete
  sample, objects with diameter $<80$kpc would fit entirely within the
  18$''$ {\it Spitzer} beam at 70\moo.

\item{\bf Flat spectrum core/jet components:} Flat spectrum core/jet
  component can also be a potential non-thermal contaminant of the
  MFIR emission. We can detect strong non-thermal jet emission at
  optical wavelengths in some objects (e.g. \object{3C273}), so it is
  reasonable to assume that this emission may contribute to the
  observed flux at MFIR wavelengths. Quasars and BLRG are known to
  have strong non-thermal beamed core components \citep{morganti97},
  with the strength of this contamination is dependent on the
  orientation of the object to the line of sight.

\end{itemize}

Identifying the non-thermal contribution to the MFIR is not a trivial
task, because the spectral shape of radio-loud AGN is often poorly
sampled in key areas such as the sub-mm and far-infrared. Although
some flat spectrum quasars and BL Lac/Blazar objects are clearly
dominated by their non-thermal component at sub-mm and infrared
wavelength, our steep spectrum selected sample has been chosen to
avoid such objects, in order to investigate the thermal emission from
dust. Despite this, the contribution from non-thermal emission from
objects with relatively strong radio cores and from steep spectrum
lobe/hotspot components in the {\it Spitzer} beam, remains a possibility, and
should be carefully investigated.

\subsection{Spectral Energy Distributions}

\begin{sidewaysfigure*}[h]
\begin{center}
\includegraphics[width =1\columnwidth]{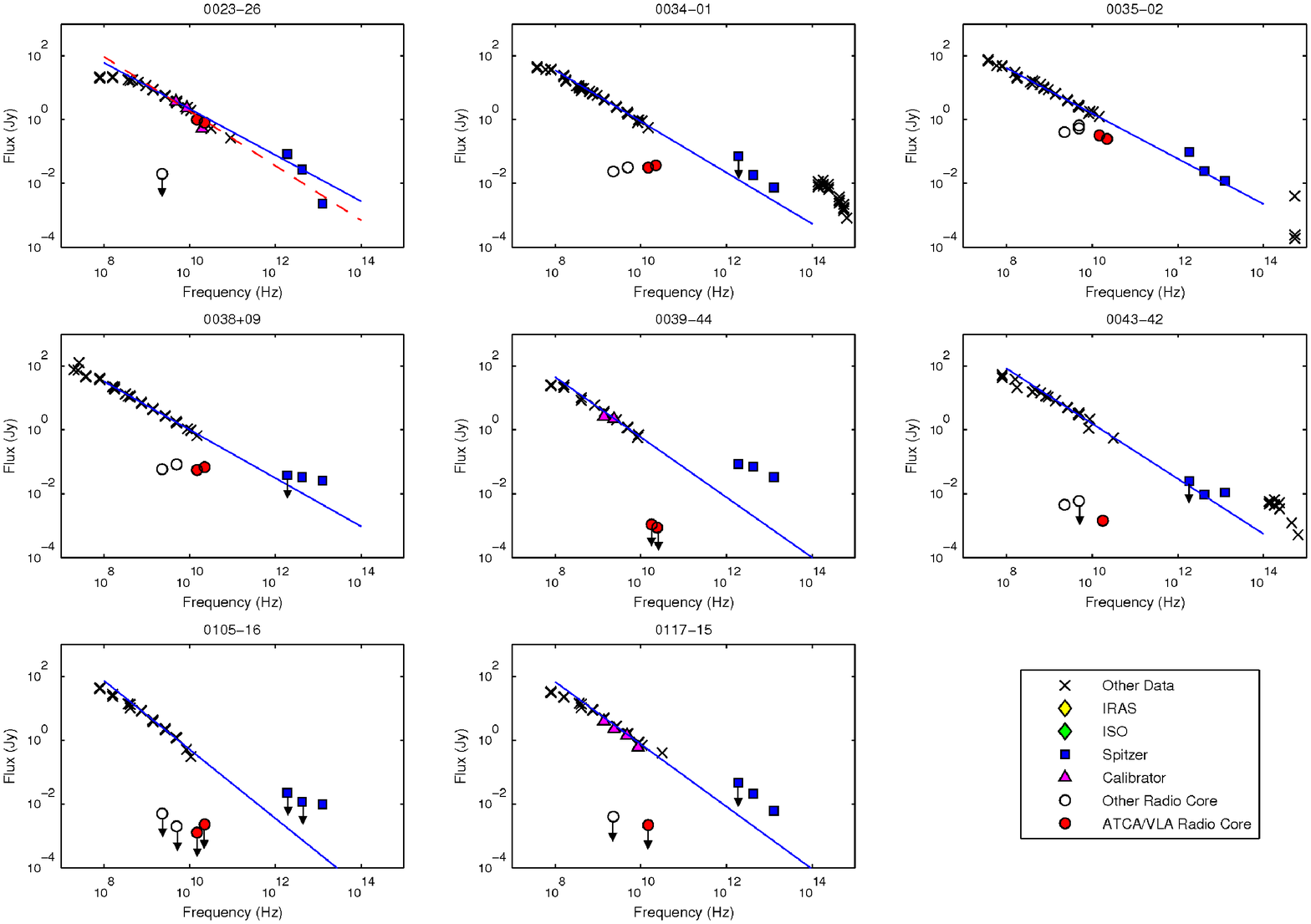}
\caption{SEDs of all objects in our sample are presented here. Upper
limits are represented by downwards pointing arrows. The blue line
is fitted to the data from $10^9$ to $10^{10}$ Hz and the red dashed
line is fitted to the data points from $10^9$ to $10^{11}$ Hz. The {\it VLA}
data in CSS object PKS0023$-$26 is the total flux measurement
and an upper limit on the core,plotted as an open circle, is taken from
\citet{tzioumis02}. \label{sed1} }
\end{center}
\end{sidewaysfigure*}

\begin{sidewaysfigure*}[h]
\begin{center}
\includegraphics[width =1\columnwidth]{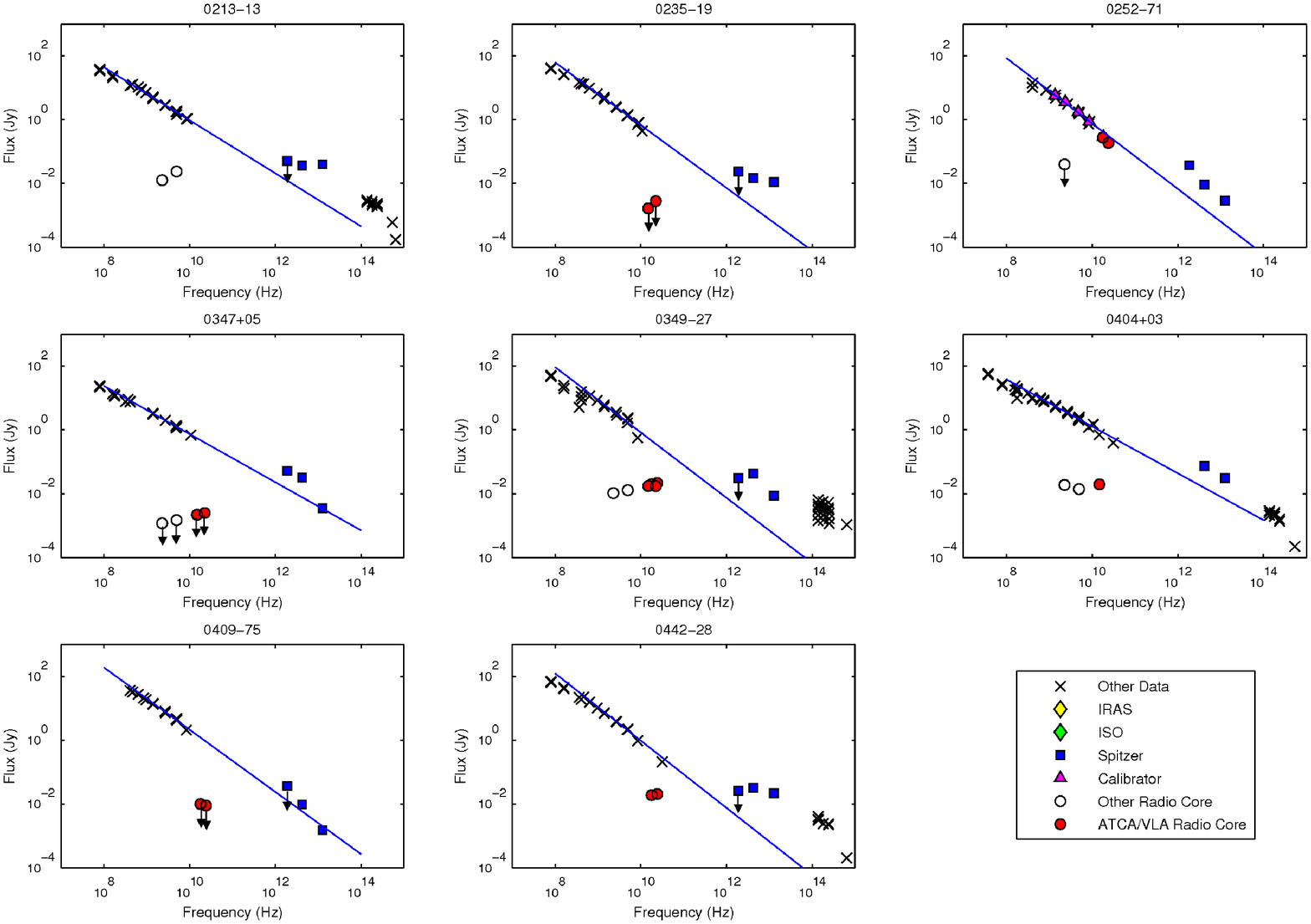}
\caption{SEDs cont. The {\it ATCA}
data in the SEDS of CSS object PKS0252$-$71 is the total flux measurement and an
upper limit on the core, plotted as an open circle, is taken from
\citet{tzioumis02}. \label{sed2} }
\end{center}
\end{sidewaysfigure*}

\begin{sidewaysfigure*}[h]
\begin{center}
\includegraphics[width =1\columnwidth]{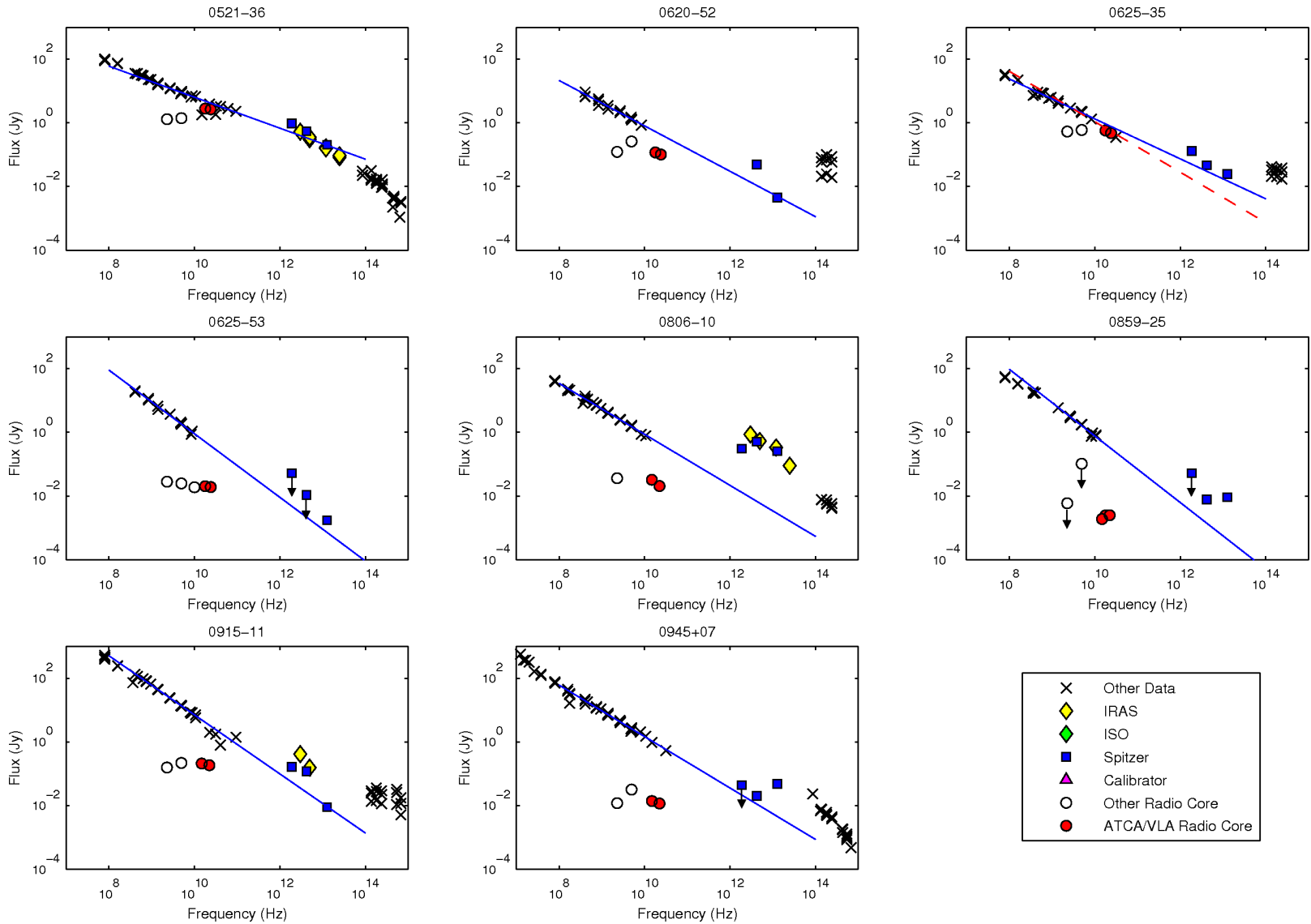}
\caption{SEDs cont.\label{sed3} }
\end{center}
\end{sidewaysfigure*}

\begin{sidewaysfigure*}[h]
\begin{center}
\includegraphics[width =1\columnwidth]{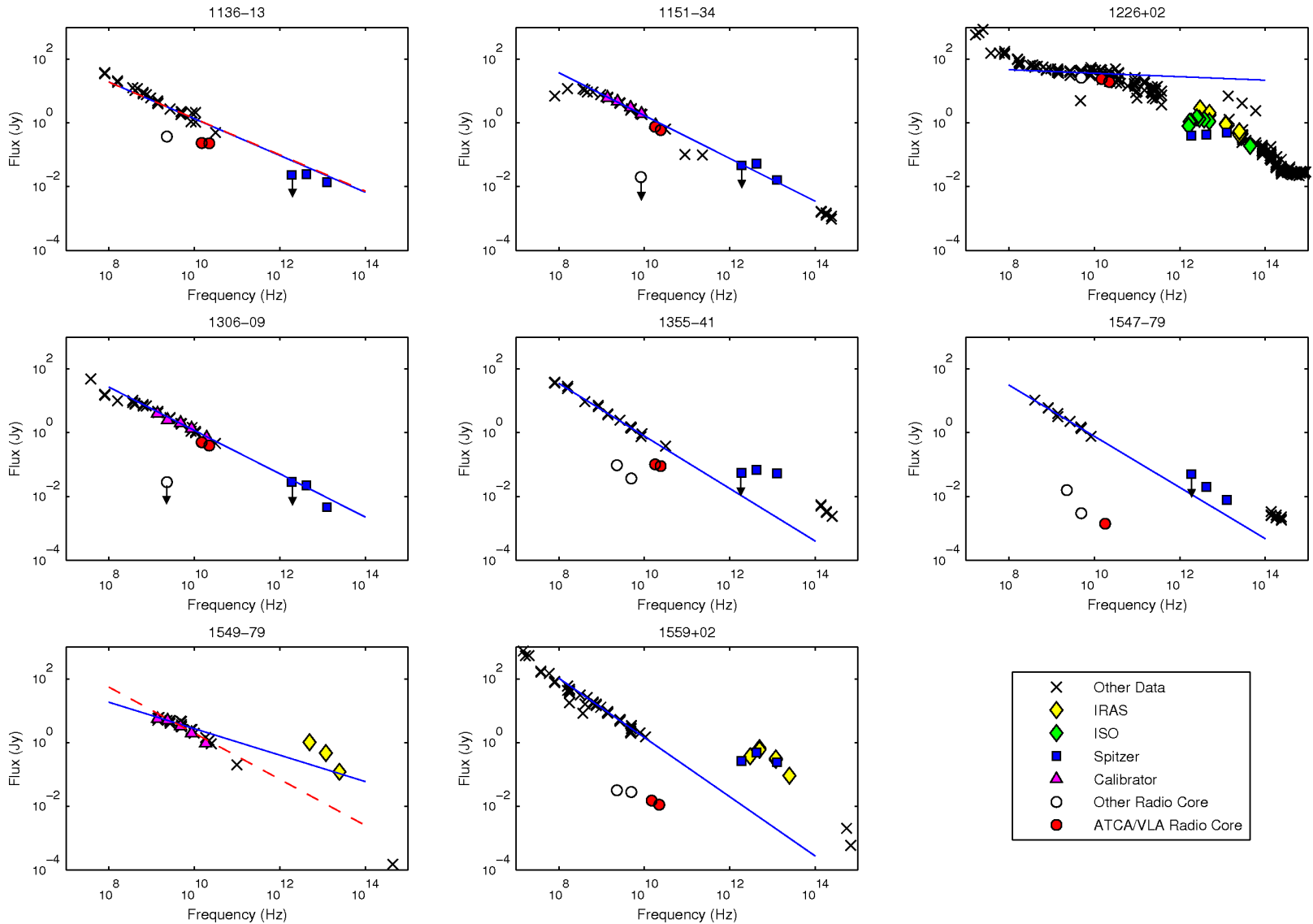}
\caption{SEDs cont. PKS1226$+$02 (3C273) clearly demonstrates its flat
spectrum dominant characteristics from the fitted power law. Note the
variability in the MFIR flux density of 3C273, where there is
approximately a decade between observations of {\it IRAS}, {\it ISO} and
{\it Spitzer}. This object is clearly in low synchrotron emission phase, see
section \ref{sec:var}.  The {\it ATCA}/{\it VLA} data in the SEDs of CSS objects
PKS1151$-$34 and PKS1306$-$09 is the total flux
measurement and an upper limits on the cores, plotted as open circles, are taken from
\citet{tzioumis02}. \label{sed4} }
\end{center}
\end{sidewaysfigure*}

\begin{sidewaysfigure*}[h]
\begin{center}
\includegraphics[width =1\columnwidth]{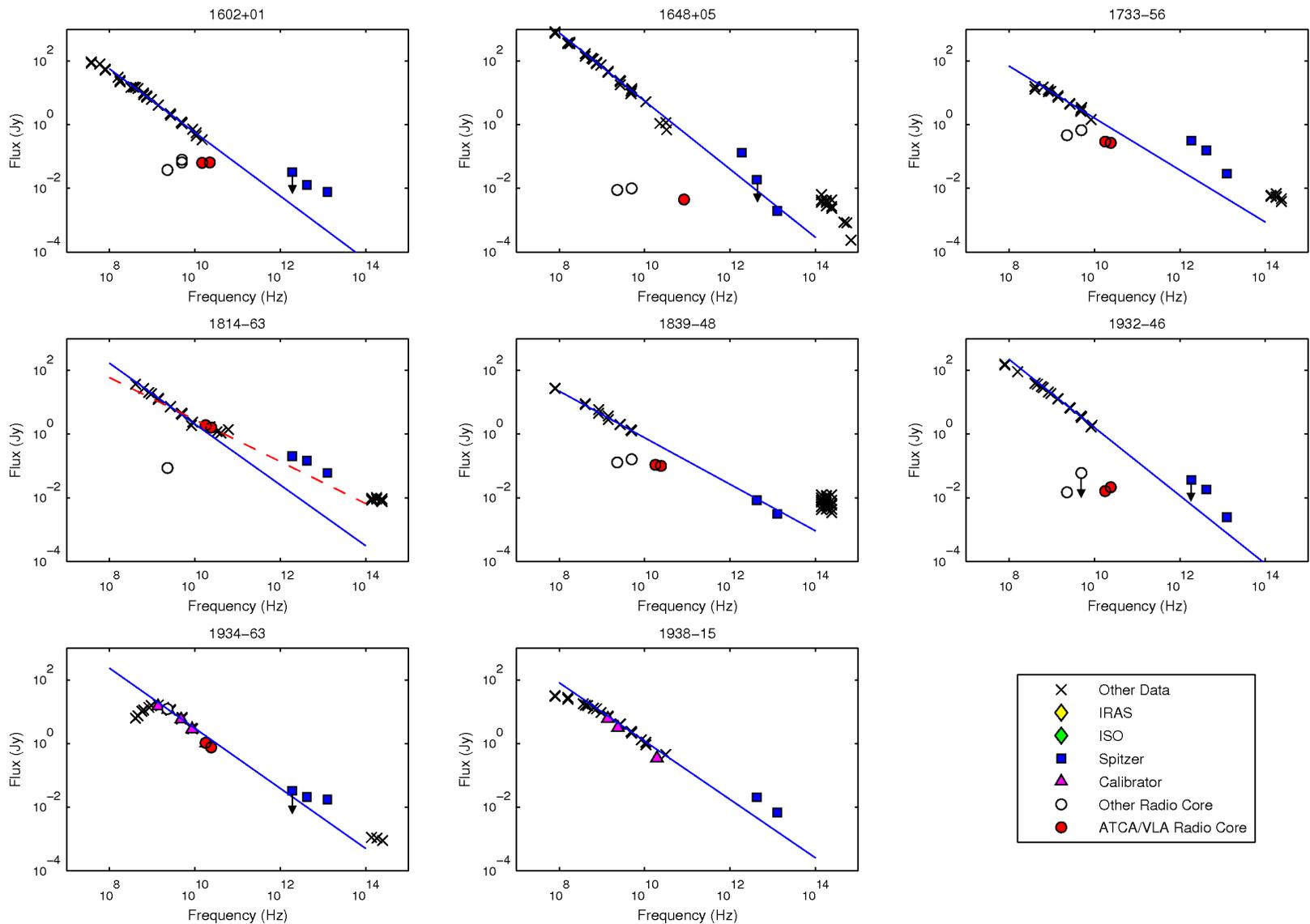}
\caption{SEDs cont. {\it ATCA}/{\it VLA} core measurement in PKS1648$+$05 from
\citet{gizani03}. PKS1938$-$15 was not observed in our high radio
frequency program. The {\it ATCA} data in the SED of CSS object PKS1814$-$63
is the total flux measurement and a possible detection of the core at
a 15 $\sigma$ level, plotted as an open circle, is taken from
\citet{tzioumis02}. We have included this PKS1938$-$15 as a candidate
for steep spectrum non-thermal contamination, despite the extrapolation
falling below the 70\moo. We believe that a jet component may be
contributing to the turn up in the high frequency radio region its
SED.\label{sed5}}
\end{center}
\end{sidewaysfigure*}

\begin{sidewaysfigure*}[h]
\begin{center}
\includegraphics[width =1\columnwidth]{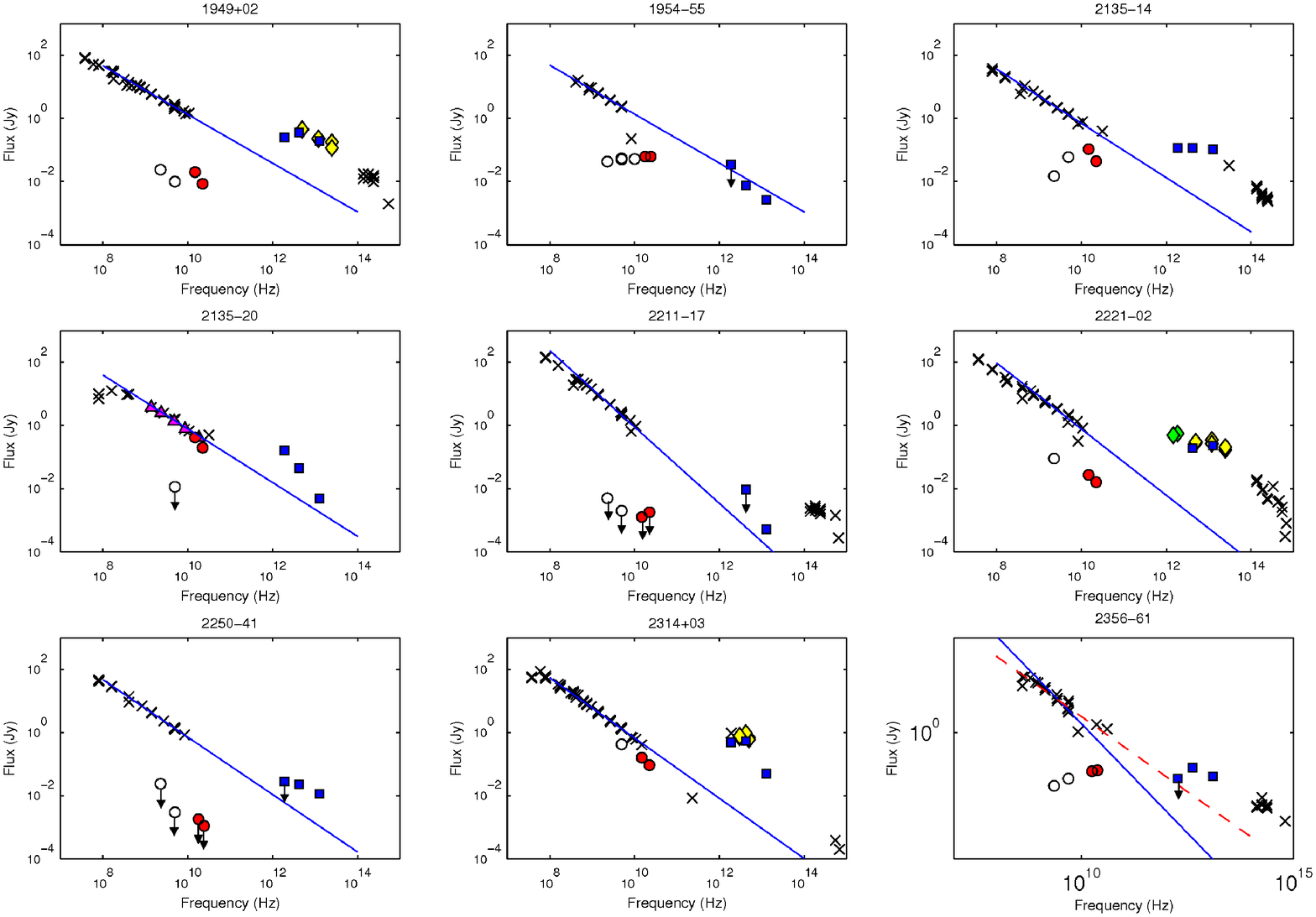}
\caption{SEDs cont. The {\it VLA}
data in the SED of CSS object PKS2135$-$20 is the total flux measurement and an
upper limit on the core, plotted as an open circle, is taken from
\citet{tzioumis02}.\label{sed6} }
\end{center}
\end{sidewaysfigure*}

In Figures \ref{sed1} to \ref{sed6} we present spectral energy
distributions for the entire sample described in section
\ref{sec:sample}. The data plotted includes the results of the
{\it Spitzer}, {\it ATCA} and {\it VLA} observations presented in this paper, along with
{\it IRAS} and {\it ISO} photometry total flux measurements and other total flux
data taken from NASA/IPAC Extragalactic Database (NED). Additional
core radio data is taken from
\citet{morganti93,morganti97,morganti99}.

Total flux observations from {\it ATCA} of a large proportion of the sample
have been made at 18.5 and 22 Ghz \citep{ricci06}, which would
complement our core measurements well. Unfortunately the atmospheric
phase stability during the latter observations was poor, which may
have contributed to the fact that, when plotted in the SEDs, many of
the values seemed unphysical compared to the other data sets. Thus
these results are not included in our SED plots.

The solid line in each plot represents a single power-law fit to the
total radio flux data of the objects between $10^9$ to $10^{10}$ Hz. A
few objects have an additional dashed line fitted to the data between
$10^9$ and $10^{11}$ Hz, when good data above $10^{10}$ Hz is
available. Unfortunately this is a relatively under-sampled region in
our data set, and we can fit this extra line to only 6 of the objects
in our sample. In the following analysis this second fitted line,
which takes better account of any high frequency steepening/flattening
of the SED, is used in preference to the first solid line fit,
where applicable.

\subsection{Lobe/hotspot synchrotron contamination}
\label{sec:lobe}
By using the fitted power laws to extrapolate the total synchrotron
radio emission through to the MFIR, it is possible to gain an
indication of whether a contribution to the MFIR flux is possible from
the non-beamed synchrotron emitting lobes and hotspots. However,
{\it Spitzer}'s beam size of 6, 18 and 40$''$ at 24, 70 and 160\moo\
respectively, means that for many targets in our sample most of the
extended steep spectrum emission lies well outside the beam. In Table
\ref{tbl-5} we address this issue, presenting a truth table from a
visual analysis of the SEDs, where we have used the 70\moo\ flux as a
reference point for the analysis.

For an object to be considered as a candidate for possible steep
spectrum synchrotron contamination, a source was required to fulfil
two criteria. Firstly, the power law extrapolation of the total (steep
spectrum dominated) radio lobe/hot-spot flux was required to fall
close to or above the 70\moo\ flux.  Secondly, a substantial fraction
of the total radio emission at 5 GHz ($>30$\%, as determined from visual inspection
of radio maps) was required to fall within the {\it Spitzer} beam at 70\moo. A tick in
columns 5 and 11 in Table \ref{tbl-5} indicates that an object fulfils
both of these criteria. We find that only 7 out of our complete sample
of 46 fulfil both these
criteria, therefore indicating that these objects have the potential
for contamination by their extended synchrotron components that fall
within the {\it Spitzer} beam. We emphasise that this is a conservative
estimate, given that the steep spectrum synchrotron component is
likely to fall quicker than the fitted power-law at higher frequencies
due to spectral aging of the electron population. Evidence for such
spectral steepening can be seen in the SEDs of \object{PKS0105$-$16},
\object{PKS0625$-$35} and \object{1549$-$79} (See Figures 5 and
6). Indeed, other authors (e.g. \citealt{shi05}, \citealt{cleary07})
fit a parabola to the steep spectrum components in order to take this
high frequency steepening into consideration. Therefore, our power-law
extrapolation is likely to provide an over-estimate of the rate of
steep spectrum contamination at MFIR wavelengths in our sample.  We
believe that more detailed parabolic or multiple power-law fits to the
total radio flux spectra are unwarranted given the lack of
sub-mm data for most of the objects.

\subsection{Core/jet synchrotron contamination}
\label{sec:core}
The thermal dust contribution to the MFIR observed flux can be
identified by extrapolating the slope of the total radio flux data to
higher frequencies and interpreting everything above that slope as the
thermal bump (e.g. \citealp{hughes97}; see previous section
\ref{sec:lobe}). However, this does not take into account any
contribution one might have from a flatter spectrum core/jet
component.

Using our high frequency radio core observations, and additional data
from the literature, we have plotted the radio core fluxes and upper limits
on the SEDs of most objects, in order to investigate their possible
contribution to the MFIR flux. Again, using the 70\moo\ flux point as
a reference and assuming the core spectral shape to be flat, we then
deem non-thermal contamination to be possible for those objects whose
high frequency radio core/jet fluxes lie level with or above the
70\moo\ flux.

Note again that this criterion is conservative in the sense that the
SEDs of flat spectrum radio core components may not remain flat up to
MFIR wavelengths as we have assumed. Indeed, the 3 flat spectrum, core
dominated objects in our sample (\object{3C273}, \object{PKS0521$-$36}
and \object{PKS1549$-$79}) all show significant declines between the
radio and the sub-mm/MFIR. In the case of \object{3C273} this decline
is ~2 orders of magnitude between the radio and MFIR.

 We have also investigated the alternative of fitting a power-law to
the core radio data and extrapolating this through to infrared
frequencies. Unfortunately, due to the small number of core data points
and relatively narrow wavelength range, there is relatively large
uncertainty in the extrapolations of the core data to the MFIR.
Future observations in the sub-mm region for the complete sample will
allow us to constrain the possible contribution of the flat spectrum
core-jet components to the MFIR more accurately.

 In columns 6 and 12 of Table \ref{tbl-5} we
present the results, finding that 11 out of 46 objects in our complete
sample have a possibility of contamination of their MFIR emission from
flat spectrum core/jet components.

\subsection{Level of potential non-thermal contamination}
\label{sec:non}

Overall, we have found that 15\% of our complete sample have potential
contamination from the steep spectrum components (lobes/hotspots), and
24\% have potential contamination from a core/jet component. Out of
the 46 objects in our full complete sample, a maximum of 30\% have the possibility
of significant non-thermal contamination of their thermal MFIR
emission from either steep spectrum components and/or a core/jet
component. However, as discussed above (\S \ref{sec:lobe} \& \S
\ref{sec:core}) these are likely to be conservative estimates, because
the strength of both the flat core and steep spectrum components
conceivably declines towards MFIR wavelengths faster than our simple
analysis assumes.

We stress to the reader that we have deliberately chosen a
conservative approach to estimating the non-thermal contamination in
this study. The results we present here are therefore likely to
represent an upper limit to the true degree of non-thermal
contamination in our sample.

\subsection{Variability}
\label{sec:var}

The two core-dominated objects \object{3C273} and
\object{PKS0521$-$36}, observed by {\it Spitzer}, were included in our
observations and this study for comparative purposes, because of the
large amount of previous data available. It is possible to identify
their non-thermal core/jet component contribution at MFIR wavelengths
if a relatively short timescale flux variation is observed, since the
synchrotron beamed component is thought to originate from a very
compact region.

The MFIR emission from 3C273 has been noted to be variable, decreasing
between {\it IRAS} (1983) and {\it ISO} (1995-98) observations in a way that is
consistent with monitoring at other wavelengths
\citep{meisenheimer01}. We can report that 3C273 has declined further
still in MFIR emission since these previous observations, by a similar
factor $\times$2 between the {\it IRAS} and {\it ISO} data (see Figure
\ref{sed4}). Because of the time scale of this variation, it is most
likely due to a change in core synchrotron emission, since the cool
extended dust is unlikely to vary on the current observing timescales
of decades. This implies that, claimed {\it ISO} detections of thermal emission from
dust underlying the powerful non-thermal emission in 3C273, are unlikely to
have been correct. The far-infrared emission now lies well below that
of the supposed thermal bump. Our data provide no evidence for the
detection of the thermal component in the MFIR SED of 3C273.

The MFIR emission of the BL Lac object \object{PKS0521-36} is also known to be dominated by
its non-thermal component, however we do not see such strong evidence
for variability.

\subsection{Comparison with previous studies}
\label{sec:notes-non}
Previous MFIR investigations have considered the relative
contributions of thermal and non-thermal emission in radio galaxies
and quasars (\citealp{hes95}, \citealp{vanbemmel98},
\citealp{polletta00} \citealp{cleary07}, \citealp{shi05}).  Studies
such as those by \citet{polletta00}, \citet{cleary07} and
\citet{shi05} fit radio to MFIR SEDs using various synchrotron and
thermal infrared emission components. Our findings agree with the
main conclusions from these studies that the proportion of objects
with non-thermal contamination of the MFIR by synchrotron emitting
components is small for an unbiased sample of radio-loud
AGN.

In the context of unified schemes, we can also investigate the optical
classifications of those objects we believe have a possibility for
core non-thermal contamination of the MFIR thermal emission. Out of
the 11 objects with possible non-thermal contamination from the cores,
6 are classified as WLRG and 5 as BLRG/Quasars. A large proportion of
BLRG/Quasars with non-thermal contamination is what we might expect if
these objects have a beamed component orientated close to the line of sight.
Overall our results our consistent with unified schemes that require a
beamed component for the BLRG/Quasar objects (e.g. \citealp{barthel89}).

For the WLRG -- some of which are FRI galaxies -- significant
non-thermal contamination is consistent with the relatively stronger
cores detected in FRI sources in general, as well as the detection of
non-thermal cores in such sources at optical/IR wavelengths
(\citealp{capetti07}, \citealp{chiaberge99}).

\section{Summary}
We have presented new MFIR and high frequency radio core data for a
complete sample of powerful southern radio galaxies. We detect
objects down to ~0.5mJy at 24\moo, ~8.4mJy at 70\moo\ and ~83.6mJy at
160\moo, obtaining a uniquely high detection rate at 24 and 70\moo. 

In addition, we have presented new high frequency radio core flux
measurements, detecting radio cores for 59\% of our complete
sample. With these data we have made a conservative estimate of the
non-thermal contribution to the MFIR continuum. Careful analysis of
the SEDs for our entire sample reveals that non-thermal contamination of the MFIR
is possible for a maximum of 30\% of the sources in our sample.

An in-depth analysis of these data will be presented in a second
paper (Dicken et al. 2008, in preparation).

\acknowledgments This work is based (in part) on observations made
with the {\it Spitzer} Space Telescope, which is operated by the Jet
Propulsion Laboratory, California Institute of Technology under a
contract with NASA. This research has made use of the NASA/IPAC
Extragalactic Database (NED) which is operated by the Jet Propulsion
Laboratory, California Institute of Technology, under contract with
the National Aeronautics and Space Administration.

D.D., acknowledges support from the STFC.

{\it Facilities:} \facility{{\it Spitzer} ({\it MIPS})}, \facility{{\it ATCA}}, \facility{{\it VLA}}.

\bibliographystyle{apj.bst} \bibliography{bib_list}

\end{document}